\documentclass[10pt,letterpaper]{article}
\usepackage[top=0.85in,left=2in,footskip=0.75in]{geometry}

\usepackage{amsmath,amssymb}

\usepackage{changepage}

\usepackage[utf8x]{inputenc}

\usepackage{textcomp,marvosym}

\usepackage{cite}

\usepackage{nameref,hyperref}

\usepackage{microtype}
\DisableLigatures[f]{encoding = *, family = * }

\usepackage[table]{xcolor}

\usepackage{array}

\newcolumntype{+}{!{\vrule width 2pt}}

\newlength\savedwidth

\raggedright
\makeatletter
\renewcommand{\@biblabel}[1]{\quad#1.}
\makeatother

\date{}

\usepackage{lastpage,fancyhdr,graphicx}
\usepackage{epstopdf}
\fancyheadoffset[L]{2.25in}
\fancyfootoffset[L]{2.25in}
\newcommand{\Y}[0]{\mathcal{Y}}

\newcommand{\vecx}[0]{\mathbf{x}}
\newcommand{\vecy}[0]{\mathbf{y}}
\newcommand{\vecz}[0]{\mathbf{z}}
\newcommand{\vecf}[0]{\mathbf{f}}
\newcommand{\vecg}[0]{\mathbf{g}}

\newcommand{\vecn}[0]{\mathbf{n}}

\newcommand{\ev}[1]{\langle #1 \rangle}

\newcommand{\vecom}[0]{\boldsymbol{\omega}}
\newcommand{\vecnu}[0]{\boldsymbol{\nu}}
\newcommand{\cov}[0]{\text{cov}}

\usepackage{caption}
\usepackage{subcaption}
\graphicspath{{Figures/}}
\usepackage{import}

\begin{document}
\vspace*{0.2in}

\begin{flushleft}
{\Large
\textbf\newline{The Neural Particle Filter} % Please use "title case" (capitalize all terms in the title except conjunctions, prepositions, and articles).
}
\newline
\\
Anna Kutschireiter\textsuperscript{*1} %\Yinyang},
Simone Carlo Surace\textsuperscript{1} %\Yinyang},
Henning Sprekeler\textsuperscript{2} %\ddag},
Jean-Pascal Pfister\textsuperscript{1} %\ddag*},
\\
\bigskip
\textbf{1} Institute of Neuroinformatics, University of Zurich and ETH Zurich, Zurich, Switzerland
\\
\textbf{2} Institute of Software Engineering and Theoretical Computer Science, Technische Universit{\"a}t Berlin, Berlin, Germany
\bigskip

% Insert additional author notes using the symbols described below. Insert symbol callouts after author names as necessary.
% 
% Remove or comment out the author notes below if they aren't used.
%
% Primary Equal Contribution Note
%\Yinyang These authors contributed equally to this work.
%
%% Additional Equal Contribution Note
%% Also use this double-dagger symbol for special authorship notes, such as senior authorship.
%\ddag These authors also contributed equally to this work.

%\textcurrency Current Address: Dept/Program/Center, Institution Name, City, State, Country % change symbol to "\textcurrency a" if more than one current address note
% \textcurrency b Insert second current address 
% \textcurrency c Insert third current address

%% Deceased author note

%% Group/Consortium Author Note
%\textpilcrow Membership list can be found in the Acknowledgments section.

% Use the asterisk to denote corresponding authorship and provide email address in note below.
* annak@ini.uzh.ch

\end{flushleft}
\section*{Abstract}

% Version \today
 
The robust estimation of dynamically changing features, such as the position of prey, is one of the hallmarks of perception. 
On an abstract, algorithmic level, nonlinear Bayesian filtering, i.e. the estimation of temporally changing signals based on the history of observations, provides a mathematical framework for dynamic perception in real time.
Since the general, nonlinear filtering problem is analytically intractable, particle filters are considered among the most powerful approaches to approximating the solution numerically.
Yet, these algorithms prevalently rely on importance weights, and thus it remains an unresolved question how the brain could implement such an inference strategy with a neuronal population.
Here, we propose the Neural Particle Filter (NPF), a weight-less particle filter that can be interpreted as the neuronal dynamics of a recurrently connected neural network that receives feed-forward input from sensory neurons and represents the posterior probability distribution in terms of samples.
Specifically, this algorithm bridges the gap between the computational task of online state estimation and an implementation that allows networks of neurons in the brain to perform nonlinear Bayesian filtering. 
The model captures not only the properties of temporal and multisensory integration according to Bayesian statistics, but also allows online learning with a maximum likelihood approach.
With an example from multisensory integration, we demonstrate that the numerical performance of the model is adequate to account for both filtering and identification problems.
Due to the weightless approach, our algorithm alleviates the 'curse of dimensionality' and thus outperforms conventional, weighted particle filters in higher dimensions for a limited number of particles. 

% Please keep the Author Summary between 150 and 200 words
% Use first person. PLOS ONE authors please skip this step. 
% Author Summary not valid for PLOS ONE submissions.   
\section*{Author Summary}

Every day, our brain is facing the challenge of making sense of the rich and dynamical stream of sensory inputs.
Those inputs are often ambiguous, noisy and sometimes even conflicting.
That we are nevertheless able to make sense of our surrounding naturally points to the important question how estimates of real-world variables that led to perceptive input, e.g.~the position or the velocity of an object, are formed.
Further, it is unknown how the computational task of real-time state estimation can be implemented in a realistic neuronal architecture.
Here, we propose an algorithm, the Neural Particle Filter, that performs state estimation, in a way that captures essential properties of perception: it takes into account prior knowledge of the environment, weights different sensory modalities according to their reliability and is able to dynamically adapt to changes.
Implemented as a neuronal dynamics, the Neural Particle Filter predicts activation properties of the neurons involved in perception.

% Use "Eq" instead of "Equation" for equation citations.
\section*{Introduction}

% Bayesian Brain and Perception
During the last decade, an increasing number of studies have stated that the brain performs probabilistic inference during perceptual tasks \cite{Knill2004,doya07bayesianbrain}.
As an act of (approximate) Bayesian inference, perception relies on noisy and incomplete data that needs to be integrated across multiple sensory modalities and weighted according to sensory reliability.
In addition, perception makes use of the strong statistical regularities of objects in our environment by forming prior beliefs about the world.
Since our environment is fundamentally dynamic, the ability to adapt to changes in real time is essential for perception.
The Bayesian brain hypothesis is supported by ample experimental evidence, ranging from psychophysical findings \cite{Wolpert1995,Kording2004,Ernst2002a} to neuronal recordings \cite{Churchland2011,Churchland2010,Orban2016} that are in line with Bayesian computation.
However, most of the studies concerned with the theory of perception consider fairly simple tasks, where the observations are created either from static hidden variables \cite{Rao1999} or from hidden variables with a discrete state-space \cite{HuangRao2016}, or the underlying dynamics are considered linear \cite{Deneve2007,Makin2015}.

% Filtering
In a dynamical setting, where temporally changing signals have to be estimated online from the history of observations, Bayesian inference is commonly referred to as `filtering'. 
In general, nonlinear Bayesian filtering is a challenging task even without the imperative of a plausible implementation on a neuronal architecture.
If the prior distribution is a Gaussian and the noisy observations depend linearly on the hidden states, the inference problem is solved by the Kalman filter \cite{Kalman1960,Kalman1961}, which has received substantial attention in the signal processing community and turns out to be of increasing importance in neuroscientific phenomenological modeling, e.g.~in a sensorimotor integration task \cite{Wolpert1995} or in estimating motor disturbances from an adaptive gain \cite{Kording2007}.
Solutions for the more general nonlinear, i.e.~non-Gaussian, filtering problem \cite{Kushner1964,Wahrscheinlichkeitstheorie1969} are analytically intractable and thus have to be approximated.

% Link filtering and samples
Sampling-based approaches have proven to be a powerful tool to solve the nonlinear filtering problem numerically. 
In principle, they enable any posterior distribution to be represented with an accuracy that depends on the number of samples.
On the one hand, so called particle methods (see for instance \cite{Doucet2000,Kantas2012a}) are well suited for dynamical priors, but suffer in high dimensions due to the degeneracy of the importance weights and it is still unclear how to implement such an inference scheme in a neuronal network.
On the other hand, Langevin sampling \cite{Welling2011,MacKay2005} and related techniques, such as the `fast sampler' in \cite{Hennequin2014Nips}, provide a promising ground for a biologically plausible implementation of neural or synaptic sampling \cite{Moreno-Bote2011,Kappel2015}, but are restricted to static generative models.
%One more line about parameter learning?

% What we do and how we do it.
Following a sampling-based approach, we propose a framework for how the brain could perform filtering from noisy sensory stimuli, considering Marr's three levels \cite{MarrVisionBook}: the computational level, the algorithmic and representational level and the implementation level.
On the first level, the computational task of dynamical state estimation is set in the context of continuous-time continuous-state nonlinear filtering theory.
Motivated by this rigorous mathematical theory, we propose a weight-less particle filter, the Neural Particle Filter (NPF), that approximates the posterior at each time step by sampling from it.
This algorithm can further be tuned by maximum likelihood learning and thus allows for rigorous corrections in the algorithmic ansatz, as well as learning the model parameters.
The NPF exhibits properties that are considered crucial for perception.
On the implementation level, we interpret the NPF as a biologically plausible neuronal dynamics and identify the particle states with activities of task-specific neurons.

\section*{Results}

The results we are presenting are subdivided in two parts:
first, we will introduce the Neural Particle Filter as a conceptual result.
This first part will cover the first two of Marr's three levels, namely i.) the computational level with a generative model layout and a task description, and ii.) the algorithmic level, which outlines our choice of representation and the approximate solution to the nonlinear filtering problem that is based on this representation.
In the second part, we demonstrate key properties of the NPF, and we we illustrate how they might serve as a model for a neuronal dynamics involved in perception. 

\subsection*{Model}
\paragraph*{Nonlinear filtering as a generic computational task}

% Hidden state
We formulate the computational task in terms of the classical filtering problem (according to standard literature on nonlinear filtering, e.g.~\cite{Jazwinski,Bain2009}).
The hidden state\footnote{For consistency, vectors will be printed in bold face, i.e.~$ \mathbf{v} = (v_{1},v_{2},\dots)^T $.} $ \vecx_t \in \mathbb{R}^n $, i.e.~the real-world variable that the brain cannot access directly, follows the It\^{o} stochastic differential equation (SDE):
\begin{eqnarray}
d\vecx_t &=& \vecf(\vecx_t)\,dt+\Sigma_x^{1/2}\, d\vecom_t, \label{eq:GenerativeModel Hidden}
\end{eqnarray}
with a nonlinear, deterministic drift function $ \vecf(\vecx):\mathbb{R}^{n} \to \mathbb{R}^{n} $.
Stochastic diffusion is governed by the uncorrelated Brownian motion process\footnote{For Brownian motion processes: $\langle d\vecom_{t}d\vecom_{s}^{T}\rangle=\mathbb{I}^{n\times n}\,dt$ if $ t=s $, otherwise $\langle d\vecom_{t}d\vecom_{s}^{T}\rangle=0$.} $\vecom_{t}\in\mathbb{R}^{n}$ with noise covariance  $ \Sigma_x \in \mathbb{R}^{n} \times \mathbb{R}^{n}$.

% Observations
At each moment in time, the hidden state $ \vecx_t $ gives rise to noisy observations $ \vecy_s \in \mathbb{R}^m $ that represent sensory input.
The observation dynamics is again modeled in terms of an It\^{o} diffusion, with a drift term following the hidden states via a generative function $ \vecg(\vecx):\mathbb{R}^{n} \to \mathbb{R}^{m} $ and a Brownian motion diffusion, modulated by the sensory noise covariance $ \Sigma_y \in \mathbb{R}^{m} \times \mathbb{R}^{m}$:
\begin{eqnarray}
	d\vecy_t &=& \vecg(\vecx_t)\,dt+\Sigma_y^{1/2}\,d\vecnu_t. \label{eq:GenerativeModel Observations}
\end{eqnarray}
Together, Eqs.~\eqref{eq:GenerativeModel Hidden} and \eqref{eq:GenerativeModel Observations} define a generative model. % (Fig.~\ref{FIG:GenerativeModel}).

% What is filtering?
Solving the filtering problem is the task of finding the posterior probability $ p(\vecx_t | \Y_t) $ of the hidden state, conditioned on the whole sequence of observations $ \Y_t = \{ \vecy_s ,s\in [0,t] \} $ up to time $ t $.
For a linear hidden dynamics $ \vecf(\vecx) $ and a linear observation dynamics $ \vecg(\vecx) $, this task is solved by the Kalman-Bucy filter \cite{Kalman1961}, which is a continuous-time version of the well-known Kalman filter.
However, the solution to the nonlinear filtering problem is in general analytically intractable, because it suffers from the so-called closure problem (see \nameref{S1}).
Therefore, introducing a suitable approximation is an inevitable step when approaching the nonlinear filtering problem.

\paragraph*{Sampling-based representation}

We approximate probability distributions in terms of a finite number of variables.
For example, this can be achieved by taking $ N $ weighted samples: % (cf.~Fig.~\ref{FIG:PPCvsSampling}B):
\begin{eqnarray}
	p(\vecx,t) &\approx& \sum_{k=1}^N w_k \, \delta(\vecx-\vecx_t^{(k)}), \ \text{with } \sum_{k=1}^N w_k = 1. \label{eq:Sampling} 
\end{eqnarray}
Thus, the probability of the random variable to have a certain value range is proportional to the relative number of samples within this range, weighted by their respective weight $ w_i $.

Filtering algorithms representing the posterior in this sampling-based manner are commonly referred to as particle filters.
In standard particle filters (such as outlined in \cite{Doucet2009}), update rules for the trajectories $ \vecx_t^{(k)} $, as well as the weights $ w_k $ are given.
Despite asymptotic convergence to the true posterior for an infinite number of particles, this approach has two disadvantages:
First, one finds numerically that after a finite number of time-steps most particle weights decay to zero, which depletes the number of effective samples.
Weight decay, or degeneration, is an undesirable trait of weighted particle methods in general.
As stated in the convergence theorem \cite[Theorem 23.5]{OxfordHBofNLFiltering2011}, the upper bound of the divergence between true posterior and the posterior estimated by the weighted particle system is a function of time, and hence might be growing due to the weight decay.
Second, the problem is exacerbated if the number of dimensions of the hidden state $ \vecx_t $ is large.
In this case, the number of particles needed for good numerical performance grows exponentially with the number of dimension, a variant of the `curse of dimensionality' \cite{Daum2003}.

% motivate this choice of representation
In the theoretical neuroscience literature, sampling-based approaches for filtering with a representation of the posterior as in Eq.~\eqref{eq:Sampling} have not received much attention so far (one of the few examples can be found in \cite{Greaves-Tunnell2015}), although they have some experimental support \cite{Berkes2011a,Churchland2010} and are considered relevant according to the neural sampling hypothesis \cite{Fiser2010}.
Therefore, we would like to explore this approach further.
To overcome the difficulties encountered with weighted approaches, we consider a particle filter with equally weighted samples, i.e.~$ w_k = 1/N \  \forall k $.

\paragraph*{Filtering with the Neural Particle Filter}

% introduce algorithm
As an inference algorithm, we propose an SDE that governs the dynamics of particles $ \vecz_t $.
Let us consider $ N $ i.i.d.~stochastic processes $ \vecz_t^{(k)},\ k=1,\dots,N $, conditioned on the observations $ \Y_t $, following the It\^{o} diffusion
\begin{eqnarray}
d \vecz_t^{(k)} &=& \vecf(\vecz_t^{(k)})\,dt + W_t \big( d\vecy_t - \vecg(\vecz_t^{(k)})\,dt \big) + \Sigma_x^{1/2} \, d\mathbf{w}_t, \label{eq:NeuralFilter}
\end{eqnarray}
where $\mathbf{w}_{t}\in\mathbb{R}^{n}$ is an uncorrelated vector Brownian motion process and $ W_t $ is a time-dependent gain matrix or decoding weight matrix.

% set in context of classical filtering algorithms (intuition for mathematicians)
Equation \eqref{eq:NeuralFilter}, which we will further refer to as the Neural Particle Filter (NPF)\footnote{In the consecutive section, we will identify the particles with neuronal activities, which is why we call it \emph{Neural} Particle Filter. Though the name is similar, our NPF is not to be confused with the `neural filtering' approach in \cite{Lo2004}, which is an unsupervised learning algorithm in an artificial neural network.}, is an ansatz that serves as a sampling-based approximation to the nonlinear filtering problem:
we consider each of the $ N $ stochastic processes $\vecz_t^{(k)} $ as an independent sample, or particle, of the true posterior $ p(\vecx_t|\Y_t) $ at every time $ t $. 
Thus, expectations from the posterior are computed according to $ \mathbb{E}[\phi(\vecx_t)|\Y_t] \approx \ev{\phi(\vecx_t)} = 1/N \sum_k \phi(\vecz_t^{(k)}) $.

This ansatz is motivated by the formal solution to the filtering problem, more precisely by the dynamics of the first posterior moment\footnote{See \nameref{S1} for an outline of the formal solution and the dynamics of the first posterior moment.\label{footSI}} and shares some important properties with classical filtering methods:
First, it is governed by both the dynamics of the hidden process $ \vecx_t $ and by a correction proportional to the so-called innovation term $ d\vecn_t = d\vecy_t - \vecg(\vecz_t)\,dt $.
The innovation term compares the sensory input $ d \vecy_t $ with the current prediction $ \vecg(\vecz_t)\,dt $ according to the single particle position, and thus can be seen as a predictive error signal \cite{Rao1999}.
Second, the gain matrix $ W_t $ determines the emphasis that is laid on new information via observations $ d\vecy_t $.
This is conceptually similar to a Kalman gain \cite{Kalman1960,Kalman1961} for a linear model, and adjusts according to the reliability of a single or multiple observations.

% set in context to Bayesian inference (intuition for non mathematicians)
The gain introduces a weighting between the prior probability distribution $ p(\vecx_t) $ induced by Eq.~\eqref{eq:GenerativeModel Hidden}, and the likelihood function $ p(\vecy_t | \vecx_t) $ induced by Eq.~\eqref{eq:GenerativeModel Observations} and thus serves as a measure for the peakedness of the likelihood.
If the decoding weight is large, the dynamics in Eq.~\eqref{eq:NeuralFilter} will entirely be determined by the innovation term, and the inter-particle variability governed by the diffusion term will be negligible.
Therefore, the resulting probability distribution is given by $ p(\vecx_t | \Y_t) \approx p(\vecz_t | \Y_t) \sim \delta(\vecz_t - \vecg^{-1} ( \frac{d\vecy_t}{dt} ) ) $.
In this limit, the deterministic observation limit, a single sample from Eq.~\eqref{eq:NeuralFilter} suffices to represent the posterior.
On the other hand, if the decoding weight is zero, new information is disregarded, and each sample evolves just like an i.i.d.~process from Eq.~\eqref{eq:GenerativeModel Hidden}.\ref{footSI}
In this case, the resulting probability distribution simply equals the stationary prior distribution $ p(\vecx_t) $.

% Determining the gain
For the gain $ W_t $, we use the ansatz $ W_t = \text{cov} ( \vecx_t,\vecg(\vecx_t)^T ) \Sigma_y^{-1} $, an empirical choice motivated by the mean dynamics of the formal solution\footnote{See footnote \ref{footSI}.}.
This gain adjusts according to the observation noise $ \Sigma_y $ as well as to the spatial ambiguity as measured by the empirical, i.e.~instantaneously estimated from the particle positions, covariance between the state $ \vecx_t $ and the observation function $ \vecg(\vecx_t) $ (Eq.~\ref{eq: weight empirical} in \nameref{section:Methods}).
Although this choice is rather heuristic, it achieves a numerical performance comparable to that of a standard particle filter (PF), as demonstrated below, and is moreover straightforward to implement by empirically estimate the covariance from the particle positions.

\paragraph*{Parameter learning}

In a more general setting, model parameters of Eqs.~\eqref{eq:GenerativeModel Hidden} and \eqref{eq:GenerativeModel Observations} may not or only partially be known, and thus need to be learned online from the stream of observations $ \Y_t $.
In this case, the NPF algorithm can be extended to include a parameter update that performs an online gradient ascent on the log likelihood
\begin{eqnarray}
	L^\text{online}_t(\theta) =  \ev{\vecg (\vecx_t)}^T \Sigma_y^{-1} \, d\vecy_t - \frac{1}{2} \ev{\vecg(\vecx_t)}^T \Sigma_y^{-1} \ev{\vecg(\vecx_t)} \,dt, \label{eq:MouraMitterCost_online}
\end{eqnarray}
which in turn is computed directly from the approximated filtering distribution itself\footnote{via $ \ev{\vecg(\vecx_t) } \approx N^{-1} \sum_k \vecg (\vecz_t ^{(k)} )$, Eq.~\eqref{eq:Posterior sample estimate} in \nameref{section:Methods}}.
It can be shown that maximizing this log likelihood is equivalent to minimizing the prediction error in continuous time (see \nameref{S1}).

Further, not only the model parameters in Eqs.~\eqref{eq:GenerativeModel Hidden} and \eqref{eq:GenerativeModel Observations}, but also the decoding parameters, i.e.~components of the decoding weight or gain matrix $ W_t $, can be learned with a maximum likelihood approach, instead of setting it according to the empirical estimate from the particle positions.
This alternative corrects for the heuristic ansatz of the NPF equation \eqref{eq:NeuralFilter} by determining the decoding weights rigorously.
In fact, it can be shown that parameter learning with a maximum likelihood approach is able to make up even for a very poor filtering ansatz by setting parameters accordingly \cite{Surace2016}.

\subsection*{The Neural Particle Filter as a neuronal dynamics for perception}

% set in context of perception
In this section, we set the computational task in the context of perception and base the implementation of the algorithm on a neuronal architecture.
With a simple example, we now illustrate that our algorithm captures the following key properties of perception \cite{doya07bayesianbrain}: 1) it relies on noisy and incomplete sensory data, 2) it uses prior knowledge of the dynamic structure of the environment 3) it efficiently combines information from several sensory modalities, and 4) it can dynamically adapt to changes in the environment.

\subsubsection*{Multisensory perception as filtering}

\begin{figure}[!h]
	\includegraphics[width=\linewidth]{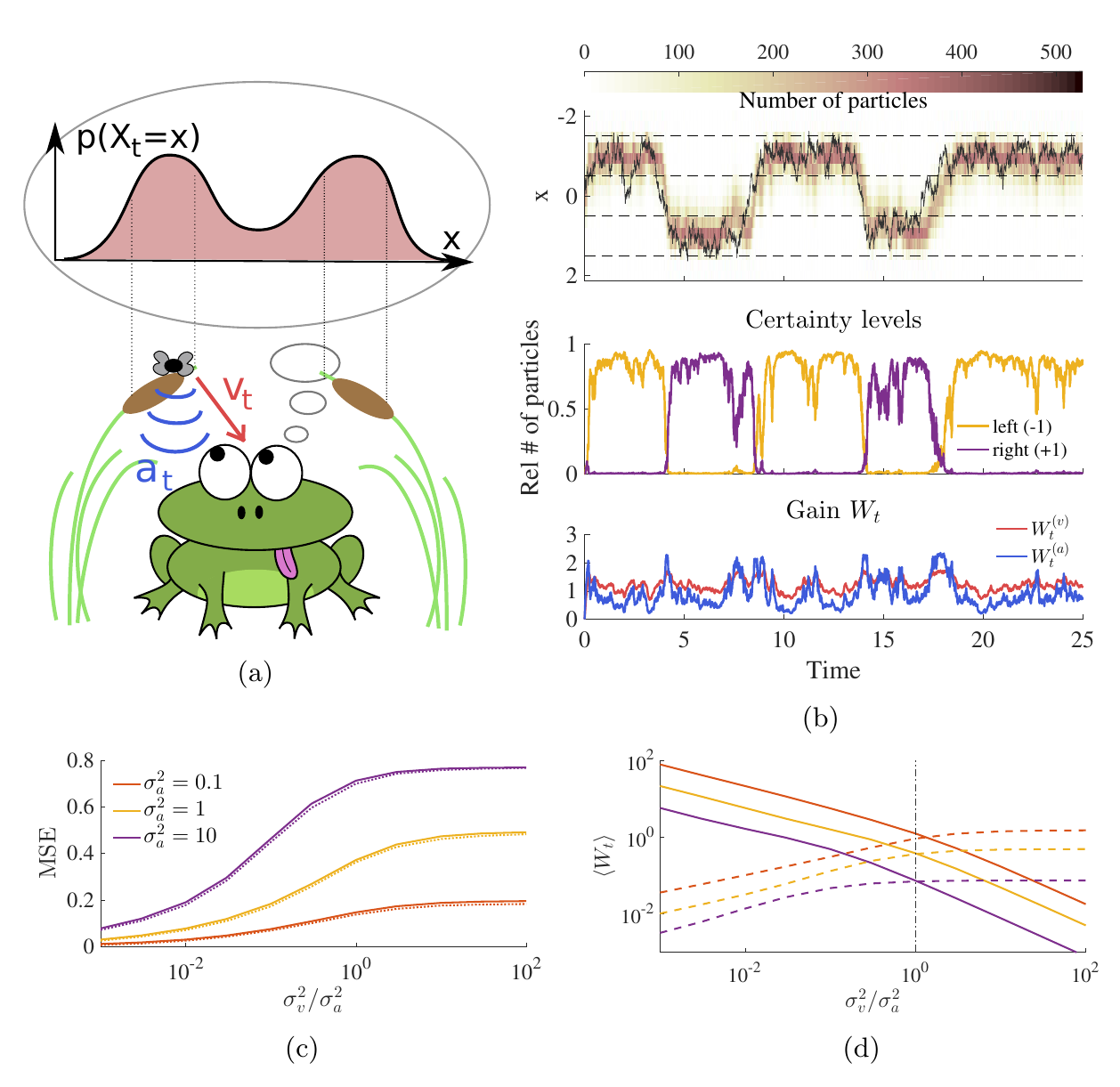}
	\caption{
		\label{FIG:Frog-and-fly}
		{\bf The NPF as a model for perception for multisensory perception.}
		{\bf{(a)}} The frog has to do filtering in order to dynamically track the position $ x_t $ of the insect, which is switching between two branches.
		We model this behavior by the stochastic dynamics in Eq.~\eqref{eq:frogfly_prior}, which gives rise to a bimodal prior distribution $ p(x_t=x) $.
		In order to track the insect, the frog has to make use of this prior, and optimally combine it with the noisy input from the sensory modalities $ v_t $ and $ a_t $.
		{\bf{(b)}} Tracking simulation with $ N=1000 $ filtering neurons and sensory noise $ \sigma_v^2=\sigma_a^2=0.1 $.
		The upper panel shows the true trajectory of the insect (solid black line) and particle densities.
		The regions between the dotted black lines denote the two branches, and certainty levels certainty levels in the middle panel correspond to the relative number of particles whose states are within one of the two branches.
		At each time, the sensory gains in the lower panel are computed according to Eqs.~\eqref{eq:frogfly_Wv} and \eqref{eq:frogfly_Wa}.
		{\bf{(c)}} Performance in terms of time-averaged mean-squared error $ MSE= \ev{(x_t-\hat{x}_t)^2}_t$ (dotted lines: standard particle filter) and 
		{\bf{(d)}} time-averaged gains $ \ev{W_t^{(v)}}_t $ (solid line) and $ \ev{W_t^{(a)}}_t $ (dashed line) as function of sensory noise.
	}
\end{figure}

Consider a frog who sits below two branches and observes an insect flying between the two branches (Fig.~\ref{FIG:Frog-and-fly}a). 
The frog wants to track the position of the insect $x_t$, which is governed by
\begin{eqnarray}
	\label{eq:frogfly_prior}
	d x_t &= &3x_t \big(1 -x_t^2 \big) \, dt+d\omega_t,
\end{eqnarray}
where the Brownian motion process $\omega_t$ accounts for noise due to the erratic behavior of the insect. 
This dynamics gives rise to a bimodal stationary distribution for the position of the insect (cf.~Fig.~\ref{FIG:Frog-and-fly}a).

The frog cannot directly observe the state $x_t$ of the insect, but instead has to rely on two sensory channels, a visual ($ v_t $) and an auditory ($ a_t $) channel.
Observation dynamics in these channels are given by
\begin{eqnarray}
\label{eq:frogfly_obsEye} dv_t &=&x_t \, dt+\sigma_v d\beta_t,\\
\label{eq:frogfly_obsEar} da_t &=&\text{tanh}(2x_t) \, dt+\sigma_a d\gamma_t,
\end{eqnarray}
where $\beta_t,\gamma_t$ are independent Brownian motions that model noise in the sensory channels, making $v_t$ and $a_t$ conditionally independent. 
The nonlinearity in the auditory channel (Eq.~\ref{eq:frogfly_obsEar}) is motivated by the fact that sound localization depends on interaural difference, which resembles a sigmoid in this example.
In order to localize the fly, the frog has to perform the task of nonlinear filtering and to compute $\hat{x}_t=\mathbb{E}[x_t|v_s,a_s,0\leq s\leq t]$, i.e.~the position of the insect, from the visual and auditory sensory streams.
Note that due to the nonlinear dynamics of the hidden and observation processes, this example is analytically intractable and thus requires an approximation.

We propose that this task is solved by a set of $N$ filtering neurons $z_t^{i}$, $i=1,...,N$.
Their neuronal dynamics are given by the NPF \eqref{eq:NeuralFilter} and for this particular example read:
\begin{eqnarray}
dz_t^{(i)} & = & 3z_t^{(i)} \big( 1-(z_t^{(i)})^2 \big) \, dt +d\omega^{(i)}_t \nonumber \\
& & +W^{(v)}_t \big(dv_t-z_t^{(i)} \, dt\big)+W^{(a)}_t\big(da_t-\text{tanh}(2z_t^{(i)}) \, dt \big), \label{eq:frogfly_NF}
\end{eqnarray}
which is governed by the dynamics of the prior as well as corrections evoked by novelty of the observations in the sensory channels, that are modulated by gains $W^{(v)}_t$ and $W^{(a)}_t$.
Thus, our model readily captures the first two key properties of perception.

The empirical distribution of neuronal activities $ z_t^{(i)} $ approximately samples the posterior distribution, thereby acting as a weight-less particle filter that successfully tracks the position of the insect (Fig.~\ref{FIG:Frog-and-fly}b).
The state estimate $\hat{x}_t$ (posterior mean) can be read out from this population by averaging the activities of the filtering neurons, i.e.~$\hat{x}_t \approx \ev{z_t} = N^{-1} \sum_i z_t^{(i)}$.

The potential of having a description of the \emph{full} posterior stretches far beyond simple state estimation, where one is only interested in the first moment.
Particularly the sampling-based approximation of this posterior allows a convenient estimation of other relevant quantities.
For example, the frog might want to know on which branch the insect is sitting in order to catch it more easily. 
The frog could directly deduce a certainty level for the left and right branch, respectively (Fig.~\ref{FIG:Frog-and-fly}b), by counting the number of neurons within a certain activity range.

\begin{figure}
	\includegraphics[width=\linewidth]{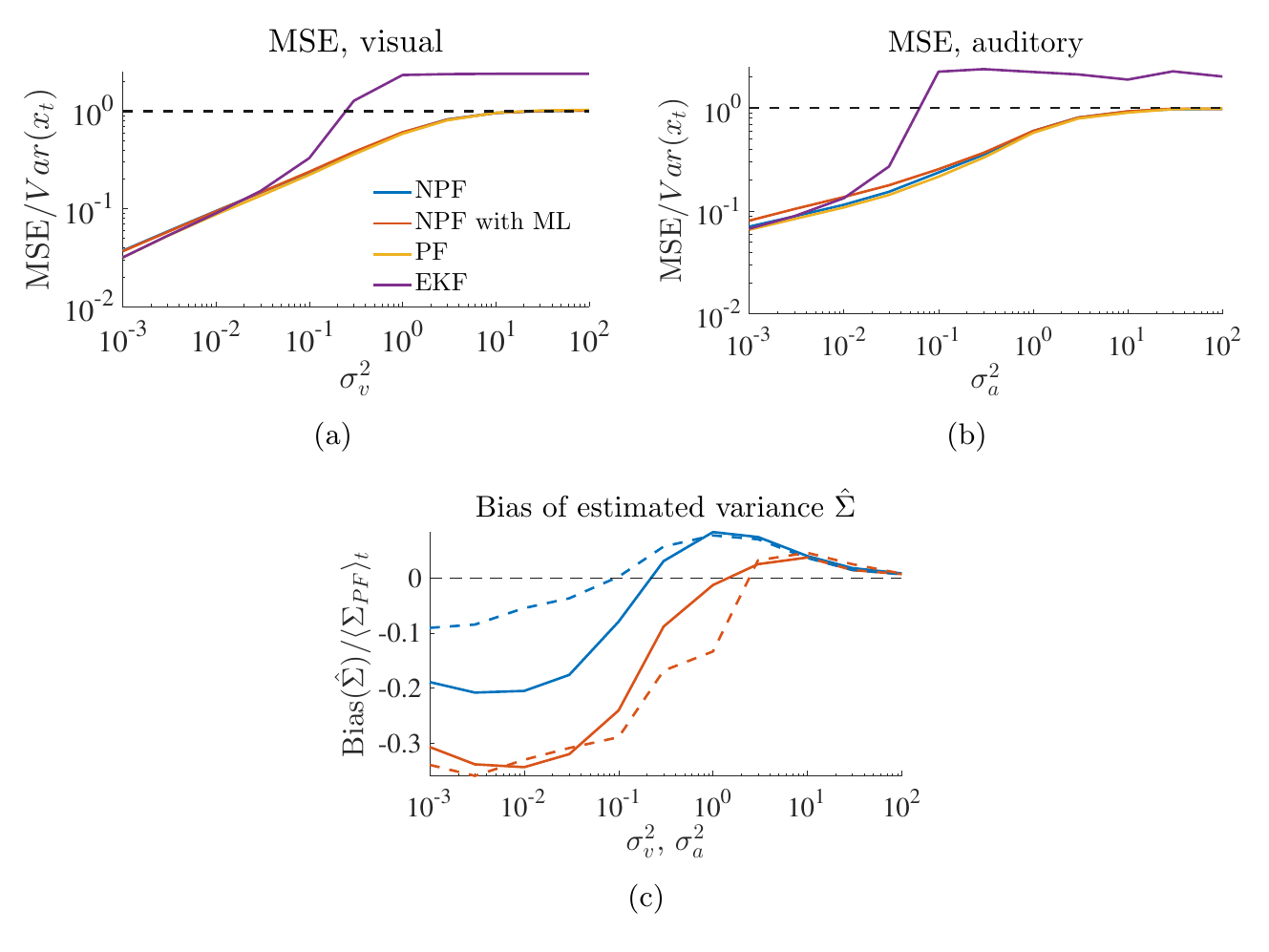}
	\caption{{\bf MSE and estimation of posterior variance from single-cue observation.}  
		Simulations correspond to the example model with only 
		{\bf (a)} visual or 
		{\bf (b)} auditory cue, i.e.~a generative model with Eqs.~\eqref{eq:frogfly_prior} and \eqref{eq:frogfly_obsEye} or \eqref{eq:frogfly_obsEar}, respectively. 
		The gain in the NPF is tuned according to $ W_t = \text{cov} ( \vecx_t,\vecg(\vecx_t)^T ) \Sigma_y^{-1} $ and according to a gradient ascent on the log likelihood (`ML'), respectively
		For benchmarking, we use a standard PF. 
		The performance of the NPF is nearly indistinguishable to that of the PF.
		In addition, we compare the performance of an EKF.
		{\bf (c)} The posterior variance $ \hat{\Sigma} $ on average exhibits a bias $ \text{Bias}(\hat{\Sigma}) = \hat{\Sigma}-\Sigma_{PF} $ when compared to that of the PF.
		Here, solid (dashed) lines refer to the visual (auditory) channel.}
	\label{FIG:Bayesian Inference}
	
\end{figure}

In a similar manner, higher-order moments of the posterior distribution can be approximated with the samples that correspond to neuronal activities.
Even though these approximated moments are not exact (Fig.~\ref{FIG:Bayesian Inference}c), the overall posterior shape is captured to a considerable extent.
For some nonlinearities, our proposed model is therefore superior to models relying on an approximation of just the first two moments of the distribution.
For instance, the Extended Kalman Filter (EKF) does by definition not capture higher order moments of the posterior and thus fails to reproduce its features.\footnote{For larger observation noise, matters become even worse because the state estimate of the EKF evolves to one of the fixed points of f(x) and remains there, irrespective of the real hidden state.} 
This explains why the predictive performance of the EKF in terms of MSE is fairly poor compared to that of the NPF (Fig.~\ref{FIG:Bayesian Inference}a,b).

\paragraph{Cue integration}
The decoding weights, or gain factors, $ W_t^{(v)} $ and $ W_t^{(a)} $, are exemplary for essential multisensory integration.
They balance the relative effects of the two sensory modalities and the prior on the dynamics of the filtering neurons and thus quantify the reliability of the sensory channels.
Here, we consider $ W_t = \text{cov}(\vecx_t,\vecg(\vecx_t)) \Sigma_y^{-1} $ in the NPF.
Thus, in this example, the weights evaluate to
\begin{eqnarray}
W_t^{(v)}=\text{Var}(x_t)\sigma_v^{-2} \label{eq:frogfly_Wv}\\
W_t^{(a)}=\text{Cov}(x_t,\text{tanh}(2x_t))\sigma_a^{-2}, \label{eq:frogfly_Wa}
\end{eqnarray}
where posterior variances and covariances are estimated empirically from the neuronal activity distribution.
The gains adjust according to both the sensory noise levels (Fig.~\ref{FIG:Frog-and-fly}d) and the spatial ambiguity evoked by the sigmoid observation function for the auditory channel $ a_t $ (Fig.~\ref{FIG:Frog-and-fly}b).
In particular, the gains become large if a channel is particularly reliable, and in extreme cases dominate the dynamics of the filtering neurons, corresponding to the deterministic observation limit discussed earlier.
The appropriate weighting of sensory information allows the neurons to solve the filtering task near-optimally and comparable to a standard PF, which is reflected by our simulation results in Fig.~\ref{FIG:Frog-and-fly}c and \ref{FIG:Bayesian Inference}a,b.

\paragraph*{Adapting to changes}

In our example, the frog could successfully track the position of the insect, but it could only do so because it had access to the generative model parameters, i.e.~it knew the prior dynamics of the insect and it it was aware of how the sensory percepts were generated from the true state of the insect.
Also, knowledge of these model parameters were crucial for determining the sensory weights $ W_t^{(v)} $ and $ W_t^{(a)} $ and thus significantly influenced the dynamics of the filtering neurons.
However, the external world, i.e.~the model parameters, does change over time, and successful perception should adapt accordingly, i.e.~the model parameters should be adjusted by online learning from the stream of observations $ \Y_t $.

\begin{figure}
	\includegraphics[width=\linewidth]{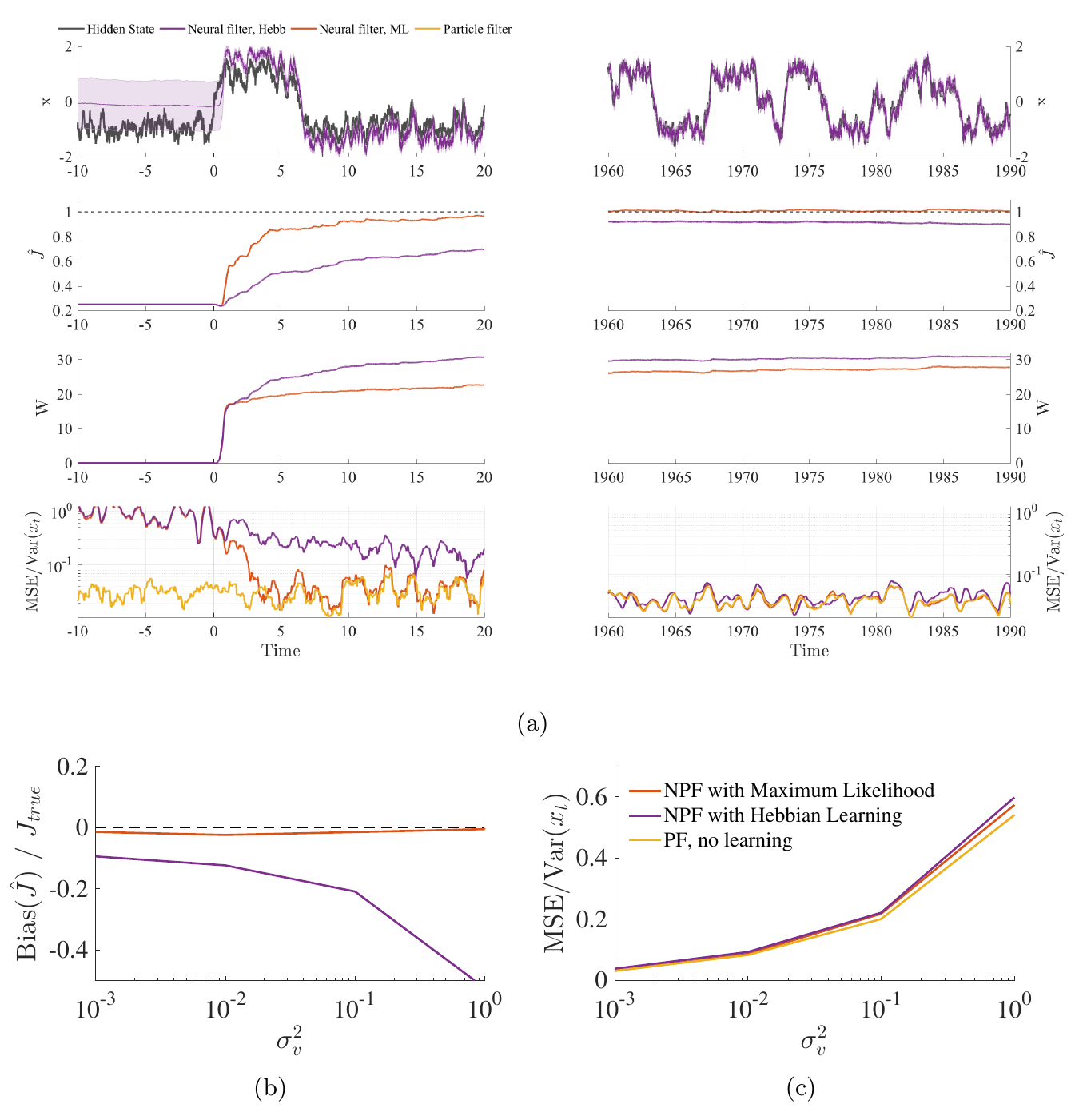}
	\caption{{\bf Model parameters are learned by a stochastic gradient ascent on the log likelihood.} 
		Simulations shown here correspond to the example model with only a visual cue, i.e.~a generative model with Eqs.~\eqref{eq:frogfly_prior} and \eqref{eq:frogfly_obsEye}. 
		The sensory gain in the equation for the filtering neurons is tuned according to maximum likelihood, the decoding weight is learned either by maximum likelihood (ML) or the Hebbian learning rule (Hebb) in Eq.~\eqref{eq:J Hebbian}, which is a valid approximation for small sensory noise. 
		As benchmark, we use a standard particle filter with the true model parameters.
		For all parameters, learning starts at $ t=0 $. 
		{\bf (a)} Filtering and online learning of the generative weight $ J=1 $ and sensory gain $ W $ for sensory noise $ \sigma_v^2=10^{-3} $. 
		As $ \hat{J} $ approaches the true value, the MSE of the NPF resembles that of the standard PF. 
		{\bf (b)} Bias of estimated generative weight $ \hat{J} $. 
		{\bf (c)} The filtering performance is not affected by the bias of $ \hat{J} $. }
	\label{FIG:Learning}
\end{figure}

%potentially redundant
We illustrate the learning of generative model parameters using our example (see \nameref{section:Methods} and \nameref{S1} for details).
This time, the frog only relies on his visual channel $ v_t $, but in addition to tracking the insect, he also has to learn the generative factor $ J $ in the function $ g(x) = J x $, which relates the position of the insect to the visual input.
Simultaneously, it also learns the gain $ W_t^{(v)} $ and with that implicitly an estimate of the reliability of its visual input.
Figure \ref{FIG:Learning} shows that this identification problem can be solved efficiently by the NPF, with an MSE that gradually approaches that of the benchmark (a standard PF with the ground-truth parameters) as the estimate of the parameters get more accurate (Fig.~\ref{FIG:Learning}a).
We find this to be true over a wide range of observation noise $ \sigma_v^2 $.
Values of the estimator $ \hat{J} $, i.e.~the learned value of the generative factor, tend to exhibit a slight negative bias, but for an observation noise of up to $ \Sigma_y = 0.1 $ still stay in a 2\%-region below the true generative weight.

It is noteworthy that the learning rule for the generative factor $ J $ can be simplified substantially for small observation noise:
\begin{eqnarray}
\eta_J^{-1}	\Delta J & \approx &  \ev{(dv_t-J z_t^{(i)} \,dt)\cdot z_t^{(i)}}. \label{eq:J Hebbian}
\end{eqnarray}
Our findings suggest that this learning rule, which we will refer to as `Hebbian' for reasons we illustrate below, leads to an estimator $ \hat{J} $ for the generative weight $ J $ which is slightly negatively biased across initial conditions (Fig.~\ref{FIG:Learning}c).
The absolute value of this bias decreases for smaller observation noise $ \Sigma_y $ and the learning rule in Eq.~\eqref{eq:J Hebbian} becomes exact for $ \Sigma_y \to 0 $.
Moreover, this bias does not seem to affect the filtering performance as measured by the MSE (Fig.~\ref{FIG:Learning}b).

\subsubsection*{Neuronal Implementation}

\paragraph{Recurrent neuronal dynamics}

\begin{figure}
	\includegraphics[width=\linewidth]{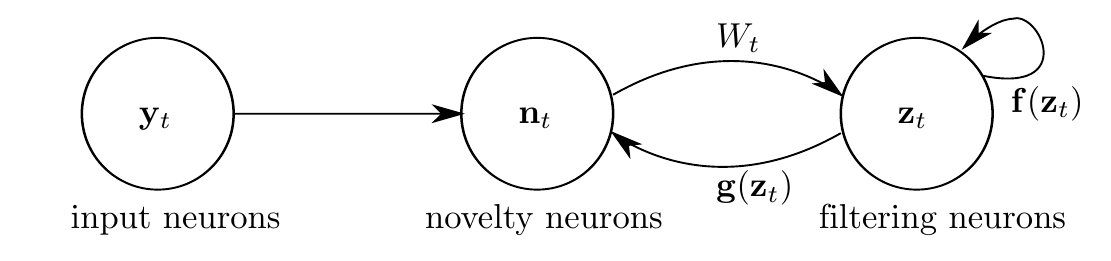}
	\caption{{\bf Neural Network Implementation}.
		Implementation of Eq.~\eqref{eq:NeuralFilter} as a recurrent neuronal network. 
		Here, $ \vecy_t $, $ \vecn_t $ and $ \vecz_t $ denote neuronal populations.}
	\label{FIG:Neural Network Implementation}
\end{figure}

In our example, we have interpreted the NPF as the neuronal dynamics of a population of $ N \times n $ filter neurons $ \vecz_t^{(i)} $, whose neuronal activities represent samples of the posterior, which is in line with the neural sampling hypothesis \cite{Fiser2010}.
Thus, analog neuronal activities are identified with the continuous particle state, for instance in terms of their instantaneous firing rate.
The internal dynamics, or self interaction, of the filtering neurons is governed by the nonlinear function $ \vecf(\vecz_t) $, which incorporates prior knowledge as a state-dependent leak.
In addition, they are recurrently connected to populations of novelty neurons $ \vecn_t $ via the feedforward synaptic weights $ W_t $ and feedback connections whose strength is governed by the nonlinearity $ \vecg(\vecz_t) $.
The dynamics of the novelty neurons are governed by the innovation term $ d\vecn_t = d\vecy_t - \vecg(\vecz_t)\,dt $.
As input to this network we consider a neuronal population $ \vecy_t $ whose rates are evoked from the underlying hidden stimulus  $ \vecx_t $ via the generative dynamics in Eq.~\eqref{eq:GenerativeModel Observations} (Fig.~\ref{FIG:Neural Network Implementation}).

\paragraph{Hebbian learning}
In this interpretation, $ W $ corresponds to the matrix of synaptic weights that connects novelty neurons $ \vecn $ to filtering neurons $ \vecz_t^{(i)} $.
If the generative function $ \vecg(\vecx_t) = J \vecx_t$ is linear, then $ J $ denotes the matrix of feedback weights which connects filtering neurons to novelty neurons.
In general, the learning rules for these weights are not local, i.e.~they rely on the state of the whole network (cf.~Eqs.~\ref{eq:dW} and \ref{eq:dJ}).
However, in the deterministic limit the learning rule for the generative weight matrix can be replaced by a learning rule that is both Hebbian and local and relies on a multiplication between pre- and postsynaptic activity (Eq.~\ref{eq:J Hebbian} and, more generally, Eq.~\ref{eq:J Learning biol rog}, see \nameref{section:Methods}).
Further, for small observation noise, $ W_t $ can be replaced by a constant matrix without affecting the filtering performance\footnote{as long as the weights are large compared to the prior dynamics}.
Therefore, at least in this limit, the network presented in Fig.~\ref{FIG:Neural Network Implementation} is implementable as a neuronal dynamics of a recurrent network with local Hebbian synaptic plasticity.

\paragraph*{The NPF alleviates the `curse of dimensionality'.}

In our example the frog had to estimate only a single hidden state, namely the one-dimensional position of the insect.
In a more realistic setting, there is a large number of hidden states, ranging from the position of an object in three-dimensional space to the relative presence of a features making up a visual scene.
Therefore, any filtering algorithm employed by a neuronal population in the brain should be economical in its resources, i.e. the number of neurons needed to solve the filtering task to a certain performance level should scale well with the number of hidden variables.
The NPF, in particular when the decoding weight $ W_t $ is determined with maximum likelihood, is able to solve the filtering task in higher dimensions with just a limited number of filtering neurons.
It thus alleviates the curse of dimensionality, which would be devastating for a realistic implementation.

\begin{figure}
	\includegraphics[width=\linewidth]{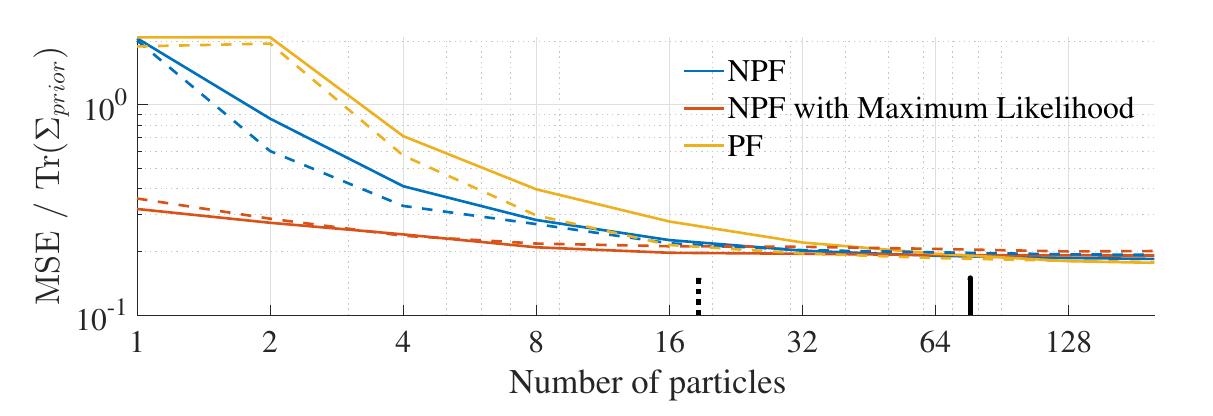}
	\caption{{\bf The NPF alleviates the `curse of dimensionality'.} 
		Toy example with nonlinear hidden dynamics and linear observation dynamics. 
		Solid lines: $ D=5 $ hidden dimensions, dashed lines $ D=1 $ hidden dimensions for comparison. 
		The error indicates the mean Euclidean distance of the estimated state $ \ev{\vecx_t} $ from the real hidden state $ \vecx_t $, normalized by the trace of the prior variance to make the error dimension-free. 
		The black marker indicates the approximate cross-over point between the Neural Filter with gain matrix $ W_t $ obtained by maximum likelihood and particle filter in $ D=5 $ (solid) and $ D=1 $ (dotted) hidden dimensions, which tends to move towards larger particle numbers as the dimensionality grows. 
		Note that the number of filtering neurons is not equal to the number of particles $ N $, but rather to $ D\cdot N $.}
	\label{FIG:PF vs NF multidim}
\end{figure}

We want to illustrate this point numerically with a toy example\footnote{see \nameref{section:Methods} for numerical details} comparing the filtering performance of the NPF in terms of its MSE for very small number of particles to that of a standard PF (Fig.~\ref{FIG:PF vs NF multidim}).
In general, we find that the NPF performs well even for a limited number of samples.
If we further allow the NPF to learn its decoding weight matrix $ W_t $ by maximum likelihood, the filter is able to solve the filtering task with a single particle almost as good as with more particles.
On the other hand, in a single-particle scenario the NPF using $ W_t = \text{cov}(\vecx_t,\vecg(\vecx_t)) \Sigma_y^{-1}  $ as well as the standard particle filter unsurprisingly exhibit an MSE, which corresponds to an independent trajectory of the prior.
Upon increasing the number of samples, the PF outperforms the NPF again, owing to the approximations employed by it. 
However, this crossover-point tends to move towards larger number of particles for higher dimensions (e.g.~Fig.~\ref{FIG:PF vs NF multidim}), which is also not surprising: 
The PF trajectories evolve according to the prior, and in higher dimensions only a few particles will be in the correct spatial domain fitting to the observation and will thus have a non-vanishing weight.\footnote{Note that this is not the weight decay due to iterative re-weighting, but rather a simple consequence of importance sampling in high dimensions, illustrating the `curse of dimensionality' for weighted particle methods.}. 

The robustness of performance for smaller number of particles, which we observe in the NPF, is mainly due to the direct influence of the observations $ d\vecy_t $ on the trajectories of the samples that correspond to the neuronal activities.
In the NPF, each neuronal activity itself can be seen as a "mini approximation" of the true posterior in terms of a $ \delta $-function, an approximation which becomes exact for very small observation noise $ \Sigma_y $.
Of course, the larger $ \Sigma_y $ becomes, the less the true posterior resembles a $ \delta $-function and the more particles are needed to account for its shape in general.

\section*{Discussion}

In this paper, we formulated the computational task of nonlinear Bayesian filtering.
Based on the theory of nonlinear filtering, we proposed an It\^{o} SDE for the posterior process and derived a learning rule for the parameters of this filter itself as well as for the generative weights of the underlying generative model.
We have thus put forward an algorithm that allows for approximate filtering in continuous time for continuous-valued hidden processes.
This algorithm allows the hidden dynamics as well as the observation dynamics to be nonlinear, and thus our model is flexible in representing a large class of general signal and emission statistics.
The sampling-based framework is a central aspect of our NPF and as such is well in line with the `Neural sampling hypothesis' \cite{Fiser2010}.

% the NF as theory of perception -> make explicit
Besides being an ansatz that is broadly consistent with neuronal dynamics, the neural filter equation we propose in Eq.~\eqref{eq:NeuralFilter} is particularly suited to model perception phenomenologically, because it shares some important properties with perception.
First, perception relies on noisy and incomplete sensory data, and uses these to make sense of the world, which in our model is reflected by inferring the hidden state variable.
Second, because prior dynamics directly enter the neuronal dynamics, prior knowledge about the environment is automatically incorporated and can in principle be learned.
Third, information from different sensory modalities is efficiently combined as a weighted input to the population of filtering neurons.
Lastly, perception can adapt to changes in the environment, which is taken into account by a dynamical gain and online parameter updates.

% Network models, following Dayan and Abbott
The implementation on a biologically realistic architecture imposes constraints on the algorithm itself as well as on how we can interpret its elements and structure.
We should always be aware that these constraints describe a highly simplified version of the real biological underpinning, but have successfully been applied in network models to qualitatively understand core computations in the brain \cite[p.~229ff]{dayan2001theoretical}.
First, neurons communicate among each other via discrete spikes, i.e.~a digital signal.
In contrast, we use the term `neuronal activity' to denote an analog quantity.
In some cases, for instance for a large number of neurons \cite[p.~231]{dayan2001theoretical}, this `analog quantity' may for instance correspond to an instantaneous firing rate.
To take into account `negative' neuronal activities, we could also consider deviations from a baseline firing rate or the membrane potential of the neuron, or the logarithm of the firing rate.
Secondly, computations in and between neurons are performed through a weighted sum of inputs from cells they are connected to, and these inputs may be a (nonlinear) transformation of the presynaptic activity.
Third, a hallmark of neural circuits is the synaptic connectivity between the neurons, the connection strength of which quantified by connective weights.
These synaptic weights are modified by learning rules, which in the most simple case are local, i.e.~they depend on the pre- and the postsynaptic neuronal activity.
The learning rules in our model are in general not local, and in fact, each filter neuron $ z^{(k)} $ has to know about the state of every other filter neuron and/or novelty neuron.
Apart from that, when parameters are learned online, it is not clear how the filter derivative should be implemented in the network.
However, we have shown that the learning rules of our model become both Hebbian and local for small observation noise, making these learning rules biologically plausible in this limit.
Lastly, because the number of neurons in the brain is finite, computations clearly have to rely on a finite number of neurons, a fact we are taking into account by representing probability distributions with samples.
Because these requirements are met by our proposed network structure, we consider our neuronal dynamics for filtering to be in line with standard network models.

\subsection*{Comparison to related work}

% Generative model
The NPF is a filtering algorithm for a continuous-time continuous-state generative model with nonlinear hidden and observation dynamics. 
Filtering algorithms based on linear generative models have been subject to extensive research. and mainly study how the analytical solution to this problem, the Kalman filter, can be implemented with neurons (e.g.~\cite{Deneve2007,Wilson2009,Makin2015,Greaves-Tunnell2015}).
However, the posterior resulting from a Kalman filter is always Gaussian, which is highly restrictive and does not properly reflect activity distributions observed in neurons (compare for instance the observations related to sparse coding as in \cite{Olshausen1996}). 
Unlike the various extensions of the Kalman filter \cite{Kalman1960} -- such as the EKF or the unscented Kalman filter \cite{Julier1997} -- which are applied to nonlinear systems, the NPF is not restricted to approximate the posterior by a Gaussian parametrized by its first and second moment. 
Rather, due to the nonlinearity in the network dynamics it may represent any probability distribution at any given time step.

% Representation
An important aspect of our work is the sampling-based representation of probability distributions, whereby the activity of each neuron is considered a single sample.
There are two main competing proposals about how the probability distributions underlying Bayesian computations might be represented in the brain.
Firstly, it has been suggested that probability distributions are expressed as probabilistic population codes (PPC \cite{Ma2006a}), in which each neuron represents a state of the encoded random variable and their activities are proportional to the probability of the corresponding state. 
Filtering approaches based on population codes, in which the neuronal activity directly relates to the posterior or the log posterior, have been explored in the literature for a large set of models, e.g.~\cite{Deneve2007,Beck2007,Sokoloski2015,Makin2015}.
In this representation, neurons directly correspond to the parameters of the distribution, and thus the critical factor for accuracy is the number of neurons.
Further, they all suffer from the `curse of dimensionality' for multimodal distribution.
The second proposal, called neural sampling hypothesis \cite{Fiser2010}, uses an inference scheme where the activity of each neuron represents a sample from the underlying probability distribution. 
Since our filtering algorithm is based on unweighted samples, our findings are in line with the advantages of the sampling-based representation outlined in \cite{Fiser2010}: it can represent any distribution without the need for a parametric form, it mitigates the `curse of dimensionality' and it is well-suited for learning.
Filtering approaches implementing Markov-chain Monte Carlo (MCMC) algorithms have received some attention lately \cite{Pecevski2011,Legenstein2014}, but since they rely on a discrete state space and assume a different coding scheme than the one suggested in \cite{Fiser2010}, the advantages listed there do not necessarily emerge from these models.

% Comparison to PF
As a filtering algorithm, the NPF is well in accordance with existing sample-based filtering approaches. 
Our ansatz may be seen as a particle filter where all particles carry the same weight and which, therefore, avoids numerical pitfalls such as weight degeneracy.
This problem is notorious in standard MCMC particle filters \cite{Doucet2000} and becomes even more severe as the number of hidden dimensions grows.
The curse of dimensionality, i.e.~the exponential growth of approximation error with the dimension of the underlying model, is an inevitable nuisance in standard MCMC approaches.
There are some tricks to deal with these problems, for instance by particle resampling or using a more refined propagator for the particles (like the optimal importance function in \cite{Doucet2000}), but neither solution is able to properly circumvent weight decay in general.
Moreover, there is currently no proposed implementation of weighted particle methods in a neural architecture.
For instance, the need to renormalize the weights at each time step introduces a coupling between the particles: While their trajectories are independent, their weights are certainly not.
On the other hand, the neural filter, not relying on importance weights in the first place, does not suffer from these numerical pitfalls and their related implementational issues.
The curse of dimensionality seems to be avoided, or at least mitigated, by the fact that the observations directly enter the particle trajectories.
However, the particles following the NPF dynamics are not completely independent either.
Coupling between the particles is mediated by the decoding weight matrix $ W_t $, whose learning rule is influenced by all the particles.
This could be avoided by fixing $ W_t $ to a constant consistent with observational noise $ \Sigma_y $, e.g.~after learning.
Even if it is numerically a bit off what we think the `real' decoding weight should be, the filtering performance is not seriously affected and particle trajectories are effectively decoupled.

% Comparison to unweighted particle filters
In the literature, there have been other approaches for particle filtering without importance weights, derived rigorously from mathematical filtering theory \cite{Yang2013,Crisan2005}.
One of these approaches, the so-called feedback particle filter \cite{Yang2013}, is based on a similar SDE for the particle trajectories as the one we propose in Eq.~\eqref{eq:NeuralFilter}.
It can be shown that the underlying distribution of the feedback particle filter evolves exactly according to the Kushner equation, whereas our approach merely approximates it.
However, the computation of the gain function in the feedback particle filter needs access to the \emph{full} filtering distribution itself, and in order to avoid numerical issues their algorithm relies on a regularization scheme that would be hard to justify biologically.
Though formally a multivariate version of the feedback particle filter exists, the gain function cannot be solved for in closed form.
The Neural Filter, though not an exact particle algorithm, overcomes this drawback by being readily applicable in higher dimensions and by being comparatively easy to compute.

% Biological plausibility
The neuronal network structure (Fig.~\ref{FIG:Neural Network Implementation}) we propose to implement the neuronal dynamics according to Eq.~\eqref{eq:NeuralFilter} is structurally similar to the one proposed in \cite{Rao1999}.
As in their model, we represent neuronal activities in terms of their instantaneous firing rate, which is an approximation to the spiking nature of biological neurons.
In their predictive coding model, a central role is assigned to the predictive error signal, which can be compared to the dynamics of the novelty neurons or novelty signal $ d \vecn_t $ in our model.
Accordingly, equations for the neuronal dynamics and for learning the generative weight in the small observation noise limit is structurally similar.
However, our model generalizes the one in \cite{Rao1999} in the sense that we allow dynamics in the prior, which is directly reflected in the dynamics of the filtering neurons.

\subsection*{Implications}

The three central aspects of our work, namely a sampling-based representation, a filtering algorithm with adaptive gain and the structure of the recurrent neuronal network, result in the following implications for the neuronal network:

\begin{enumerate}
	\item The network is robust against neuronal failure.
	\item An internal model about the world becomes manifest in internal neuronal dynamics
	\item Neuronal variability is tuned according to sensory reliability.
	\item Neurons may code for the novelty given by discrepancy between prediction by an internal model and actual observations.
\end{enumerate}

The first implication follows directly from the sampling-based representation, namely robustness against neuronal failure.
For example, if a distribution is represented by 1000 particles, removing 10\% will not significantly decrease the ability of the other particles to represent the probability distribution.
In an extreme case, we could consider a single neuron to represent the whole distribution, given that its activity state can take values in the same range as the hidden state and we allow it to sample the distribution in time.
Apart from that, as we have seen numerically, the ability to perform filtering with a reduced number of particles is not affected by particle removal to a large extent either, at least not in the particular algorithm we propose.
However, some degree of plasticity or rewiring would be necessary in order to read out expectations from the decimated neuronal population.
On the other hand, in parametric representations such as a PPC, where each neuron determines the height of a particular tuning curve assigned to it and that actually rely on the tuning curves to cover the space densely \cite{Ma2006a}, neuronal failure can be devastating:
With each neuron that breaks down, a particular point in state space cannot be represented directly anymore.
Clearly, a single neuron would never be able to account for any other distribution than the one resembling its own tuning curve.

The second and third implication is a consequence of the adaptive gain, i.e.~the decoding weight $ W_t $, that determines the emphasis that is laid on new observations.
As we have demonstrated, the gain $ W_t $ in our model increases with sensory reliability both according to sensory noise and ambiguity in the input generation,  putting more emphasis on the observations versus its internal model.
In the absence of observations, observation noise is maximal and the neuronal dynamics follow those of the hidden state which comprises an internal model about the world.
With availability of observations, sensory reliability naturally increases and variability across samples should decrease, because now their dynamics are influenced by the same stimulus via the gain $ W_t $.
Indeed, it has been found that spontaneous neuronal activities relate to prior expectations about a stimulus in visual cortex \cite{Berkes2011a}.
Further, it has been shown that inter-trial variability of neuronal responses declines upon stimulus onset \cite{Churchland2010}.
Both these experimental findings are nicely in line with our theoretical predictions.

\section*{Conclusion and Outlook}

With the Neural Particle Filter we have come up with an algorithm that allows neurons to perform nonlinear Bayesian filtering in a sampling-based manner.
Specifically, the neuronal implementation is based on a network of recurrently connected analog neurons whose dynamics are governed by the NPF algorithm.
In future work, the biological plausibility of this recurrent network model will be further addressed. 
First, observing that the learning rules in general, i.e.~for nonvanishing observation noise, do not fulfill the requirement of locality, which is needed for a biologically plausible learning rule (e.g. a Hebbian learning rule), the model could be enhanced such that individual filtering neurons obey different rather than identical dynamics.
We could for instance consider $ N $ different subnetworks and locally determine the weights in these subnetworks, possibly taking into account a (slowly-changing) global modulation factor.
Such an approach would also effectively decouple the particles on a smaller timescale.
Second, by including the theory of filtering and identification of point processes, our algorithm could be extended such that a spike-based representation may be accounted for.

\section*{Methods}
\label{section:Methods}

\subsection*{Details on numerical experiments}

\subsubsection*{The choice of the decoding weight $ W_t $}

In our numerical simulations, we consider and compare two choices of the decoding weight $ W_t $, which we will quickly motivate here.

As a first choice, we consider  $ W_t = \cov(\vecx,\vecg(\vecx)^T) \Sigma_y^{-1} $, a choice inspired by the dynamics of the formal solution.
In particular, consider the dynamics of the first posterior moment\footnote{see \nameref{S1}}, which by comparison with the NPF equation (Eq.~\ref{eq:NeuralFilter}) directly motivates this particular choice of the decoding weights.
Thereby, the covariance $ \text{cov}(\vecx,\vecg(\vecx)^T)  $ is estimated empirically from the samples $ \vecz_t^{(k)} $via
\begin{eqnarray}
\cov(\vecx,\vecg(\vecx))  &\approx & \frac{1}{N} \sum_{k=1}^{N} \vecz_t^{(k)} \vecg(\vecz_t^{(k)})^T - \frac{1}{N^2} \sum_{k,l=1}^{N} \vecz_t^{(k)} \vecg(\vecz_t^{(l)})^T. \label{eq: weight empirical}
\end{eqnarray}

As a second choice, we consider $ W = W_\text{ML} $, i.e.~the decoding weight $ W_t $ and thus the Neural Filter itself is tuned via maximum likelihood.
The learning rule is given in Eq.~\eqref{eq:dW} and corresponds to an online update of the decoding weight at each time step.
What is peculiar about this choice is that here a \emph{decoding} rather than a generative model parameter is learned, illustrating that inference and learning can be very intertwined:
In fact, it is actually possible to rephrase the filtering problem in terms of a learning problem by giving an ansatz for a decoder whose parameters have to be learned such that it can perform  filtering, but we will not consider this rather extreme case here.

\subsubsection*{Dynamics and parameters}

For our simulations, we use a nonlinear hidden dynamics, that was chosen to have a bimodal stationary distribution:
\begin{eqnarray}
d x_t &=& a x_t( b - x_t^2 ) \, dt + \sigma_x d\omega_t. \label{eq:HiddenDynamics_bimodal}
\end{eqnarray}
The parameters $ a > 0 $ and $ b > 0 $ can be used to tune the shape of the bimodal distribution, whereby the positions of the two modes is determined by $ \pm b $ and $ a $ defines how sharply the distribution is peaked around the modes.
Unless stated otherwise, parameters for the deterministic dynamics were $ a= 3 $ and $ b = 1 $, resulting in a bimodal distribution with distinct, but not too sharp peeks at $ \pm 1 $, such that it is possible for the hidden state to switch from one mode to the other.

In our examples in Fig.~\ref{FIG:Frog-and-fly}, we employ both linear and sigmoid observation dynamics $ g(x) $, thereby simulating multisensory integration with our model (cf.~Eqs.~\ref{eq:frogfly_obsEar} and \ref{eq:frogfly_obsEye}).
For the plots in Fig.~\ref{FIG:Bayesian Inference}, we use only one sensory modality per row.
The observation noise $ \sigma_v $ and $ \sigma_a $ is varied between $ 10^{-4} $ and $ 300 $.

In the multidimensional simulations in Fig.~\ref{FIG:PF vs NF multidim}, the hidden dynamics within each dimension is independent of the other dimensions and and corresponds to Eq.~\eqref{eq:HiddenDynamics_bimodal}, with $ \Sigma_x $ chosen to be the unit matrix.
The stationary distribution of this dynamics is thus a multimodal distribution with $ 2^n $ peaks for $ n $ dimensions.
The linear generative function is given by $ \vecg(\vecx) = J \vecx $ with $ J $ a rotation matrix rotating the hidden state vector by 30 degrees around each spatial axis.
In the simulations shown in Fig.~\ref{FIG:PF vs NF multidim}, $ \Sigma_y = 0.1 \cdot \mathbb{I} $.

Mean-squared errors (MSE) and biases of estimated quantities or parameters $ \hat{\theta} $ were computed by
\begin{eqnarray}
MSE &=& \frac{1}{T} \sum_{t=1}^{T} |\vecx_t - \ev{\vecx} |^2,\\
Bias(\hat{\theta}) &=& \frac{1}{T} \sum_{t=1}^{T} \hat{\theta}_t - \theta_t,
\end{eqnarray}
where $ \theta_t $ denotes the true or benchmark (particle filter) value.
Unless stated otherwise, MSEs are normalized with respect to the trace of the stationary prior variance $ \Sigma_{prior} $ to make performance comparable and, if needed, independent of the number of hidden dimensions.

Other simulation parameters comprise the time step size, which was set to $ dt=0.005 $ throughout all simulations.
All simulations were run for 500'000 time steps, corresponding to 2500 time units.
Unless stated otherwise, MSEs and biases were averaged over the last 1000 time units, equaling 200'000 time steps.

\subsubsection*{Benchmark models}

We would like to stress that the generative model we chose in our examples are nonlinear both in prior as well as in observation dynamics.
This implies that a closed form solution to this problem does \emph{not} exist, and thus approximations have to be employed.
Therefore, when assessing the performance of the NPF, we compare it to two approximate filtering algorithms, both of which widely used in approximating the posterior distribution in nonlinear filtering problems: a standard particle filter \cite{Doucet2000} and continuous-time version of the extended Kalman filter \cite{Jazwinski}.
For further information on these algorithms, see \nameref{S1}.

\subsection*{Maximum likelihood parameter learning}

\subsubsection*{The cost function}

For the continuous-time continuous state-space generative model given by Eqs.~\eqref{eq:GenerativeModel Hidden} and \eqref{eq:GenerativeModel Observations}, learning in the mathematical literature commonly referred to as `system identification', is a tough problem which has hardly been looked at.
In fact, the (to our knowledge) only reference which gives an explicit cost function for identification in our setting is a technical report by Moura and Mitter \cite{Mourat1986}.
Based on a change of probability measure, they propose a cost function that is equivalent to the log likelihood of the input $ \log p_\theta(\Y_t) $ with model parameters $ \theta $, which for our model reads:
\begin{eqnarray}
L_t^{\text{offline}}(\theta) = \int_0^t \ev{\vecg (\vecx_s)}^T \Sigma_y^{-1} \, d\vecy_s - \frac{1}{2} \ev{\vecg(\vecx_s)}^T \Sigma_y^{-1} \ev{\vecg(\vecx_s)} \,ds, \label{eq:MouraMitterCost_offline}
\end{eqnarray}
where expectations are taken with respect to the filtering distribution at time $ s $, $ p( \vecx_s | \Y_s) $. 
%\footnote{In the SI, we further offer a more intuitive justification for this cost function in Eq.~\eqref{eq:MouraMitterCost_offline} in terms of a minimization of the reconstruction error. }

In a discrete-time approximation, Eq.~\eqref{eq:MouraMitterCost_offline} immediately suggests an online maximization scheme.
Instead of maximizing the cost function at a time $ t $ for the whole observation sequence $ \Y_t $, implying we would have to run the filter all over again each time we change the parameters, we just perform a gradient ascent with respect to the parameters $ \theta $ on the last contribution to the integral, i.e.~to
\begin{equation*}
L^\text{online}_t(\theta) =  \ev{\vecg (\vecx_t)}^T \Sigma_y^{-1} \, d\vecy_t - \frac{1}{2} \ev{\vecg(\vecx_t)}^T \Sigma_y^{-1} \ev{\vecg(\vecx_t)} \,dt, \tag{\ref{eq:MouraMitterCost_online}}
\end{equation*}
where expectations are with respect to the filtering distribution at time $ t $, $ p(\vecx_t|\Y_t) $.
It can be shown that maximization of this cost function is equivalent to a minimization of the reconstruction error $ ( d \vecy_t - \vecg(\vecx_t) \,dt )^2 $ at each time step (see \nameref{S1} for a proof).

\subsubsection*{Learning rules}

Parameter learning is implemented by maximizing the log likelihood (Eq.~\ref{eq:MouraMitterCost_online}) by a gradient ascent with respect to the model parameters $ \theta $, giving rise to the following online learning rule for the parameters:
\begin{eqnarray}
\eta_\theta^{-1} \Delta \theta &=&  \Big( \frac{\partial}{\partial \theta} \ev{\vecg(\vecx_t)}\Big)^T \Sigma_y^{-1} \big(d\vecy_t -  \ev{\vecg(\vecx_t)}\, dt  \big). \label{eq:ParaLearning online}
\end{eqnarray} 
This approximation of learning is justified if the time scale of learning is much larger than the dynamics of the filter, i.e.~for small learning rates.

Equation \eqref{eq:ParaLearning online} exhibits a peculiar structure:
The novelty signal $ d\vecy_t -  \ev{\vecg(\vecx_t)}\, dt $ is multiplied with a parameter gradient on the posterior estimate of the generative function $ \ev{\vecg(\vecx_t)}$.
Thus, we have to take into account the implicit change of the posterior distribution, the so called \emph{filter derivative}, with respect to the model parameters.
The filter derivative is in general hard or even impossible to compute analytically and many identification problems deal with estimating it (cf.~for instance \cite{Kantas2012a,Surace2016}).

In our model, we make use of the approximated posterior dynamics in order to derive dynamics of the filter derivative for parameter learning.
Equation \ref{eq:ParaLearning online} can be approximated by taking $ N $ samples from the NPF equation (\ref{eq:NeuralFilter}) in order to express the posterior estimates:
\begin{eqnarray}
\ev{\vecg(\vecx_t)} &\approx& \frac{1}{N} \sum_{k=1}^N \vecg(\vecz_t^{(k)}), \\
\frac{\partial}{\partial \theta} \ev{\vecg(\vecx_t )} &\approx& \frac{1}{N} \sum_{k=1}^N G(\vecz_t^{(k)}) \frac{\partial \vecz_t^{(k)}}{\partial \theta}, \label{eq:Posterior sample estimate}
\end{eqnarray}
where $G_{ij}(\vecz^{(k)}) = \frac{\partial g_i}{\partial x_j} (\vecz^{(k)})  $ denotes the Jacobian of the generative function.
Thus, the filter derivative is taken into account by considering the infinitesimal change in the position of sample $ \vecz^{(k)} $ with respect to the change in parameter value $ \theta $.

The single particle filter derivative, $ \frac{\partial \vecx_t ^{(k)} }{\partial \theta } $, cannot be computed directly.
However, based on Eq.~\eqref{eq:NeuralFilter}, it is possible to compute its dynamics:
\begin{eqnarray}
	d \left( \frac{\partial \vecz_t ^{(k)} }{\partial \theta } \right) &=& \frac{\partial}{\partial \theta } \big( d \vecz_t^{(k)} \big).
\end{eqnarray}
Note that every single parameter that is learned has $ N $ accompanying filter derivatives of this form.

In this work we are interested in learning the decoding weight matrix $ W_t $ and, for a linear observation dynamics $ \vecg(\vecx) = J \vecx $, in learning the generative matrix J, respectively.
The resulting learning rules for the components of the decoding weight matrix with learning rate $ \eta_W $ are given by
\begin{eqnarray}
		\Delta W_{ij} & = & \eta_W  \frac{\partial}{\partial_{W_{ij}}} \ev{\vecg(\vecx_t)}^T \Sigma_y^{-1} \left( d\vecy_t - \ev{\vecg(\vecx_t) } dt \right), \label{eq:dW}
\end{eqnarray}
with filter derivative dynamics
\begin{eqnarray}
d \left( \frac{\partial \vecz_t ^{(k)} }{\partial W_{ij} }  \right) & = & \left( F( \vecz_{t}^{(k)} ) - W G(\vecz^{(k)}_t) \right) \frac{\partial \vecz_t ^{(k)} }{\partial W_{ij} } dt + \big[d \vecy_{t} - \vecg(\vecz^{(k)}_t)dt \big]_{j} \mathbf{e}_{i}, \label{eq:d dx dWij}
\end{eqnarray}
where $ F_{ij}(\vecz^{(k)}) = \frac{\partial f_i}{\partial x_j} (\vecz^{(k)})  $ denotes the Jacobian of the nonlinear hidden dynamics and  $ \mathbf{e}_{i} $ denotes the unit vector in the $ i $-th direction.
This implies that, when we take $ W_t $ to be a plastic decoding weight matrix that is learned as observations become available, at least three equations are needed to infer the hidden state at each time step:
First, Eq.~\eqref{eq:NeuralFilter} to evolve the states of the filter neurons, and second, Eqs.~\eqref{eq:dW} and \eqref{eq:d dx dWij} to update the weights in the filter equation.

Analogously, learning rules for the components of the generative matrix for linear observation dynamics $ \vecg(\vecx) = J \vecx $ read
\begin{eqnarray}
\Delta J_{ij} & = & \eta_J \left[ \left( \frac{\partial \ev{\vecx_t}}{\partial J_{ij}} \right)^T J^T \Sigma_y^{-1} (d\vecy_t - J \ev{\vecx_t} \,dt)  + \Big( \Sigma_y^{-1}  (d\vecy_t - J \ev{\vecx_t} \,dt)  \Big)_{ij} \right]. \label{eq:dJ}
\end{eqnarray}
In addition to a term proportional to the filter derivative, the learning rule contains a second term that emerges from an explicit dependence of the likelihood in the generative weight. % i.e.~$ \delta J_{ij} \propto d \ev{n_i}  \, d\ev{x_j} = (d\vecy_t - J \ev{\vecx_t}\,dt)_i \ev{\vecx_t}_j$.
Filter derivatives are given by
\begin{eqnarray}
d \left( \frac{\partial \vecz_t ^{(k)} }{\partial J_{ij} } \right) & = &  \left( F( \vecz_{t}^{(k)} ) - WJ \right) \frac{\partial \vecz_t ^{(k)} }{\partial J_{ij} } dt - \hat{x}_{t,j}^{(k)} W \mathbf{e}_{i} dt. \label{eq:dNuJ}
\end{eqnarray}

\paragraph{Bias due to sampling}
In the derivation of these learning rules, we used a sampling-based representation of the approximated posterior in order to estimate log-likelihood and gradients.
This introduces a bias in these estimations, which we should correct for when computing parameter estimates.
Unfortunately, this bias is in general not analytically accessible, but at least for a linear generative model it can be shown that it vanishes with $ 1/N $ (for a proof, see \nameref{S1}).

\paragraph{Approximation for small observation noise}
The learning rules we obtain for the decoding weights $ W_t $ and the generative weights $ J $ are not local, implying that the weights can only be computed when knowing the state of each filter neuron at each time.
However, for small observation noise the learning rule for the generative weight $ J $ can be approximated by a local learning rule with a Hebbian structure.
First, we can neglect the filter derivative, which decays to zero very fast because in this limit, the decoding weight $ W_t $ is generally large (cf.~Eq.~\ref{eq:dNuJ}), and thus the first term in Eq.~\eqref{eq:dJ} vanishes.
Second, because in this limit the posterior will approach a $ \delta $-distribution around the true hidden state $ \vecx $, as does the approximated posterior, we can approximate the learning rule for $ J $ by:
\begin{eqnarray}
\eta_J^{-1}	\Delta J & \propto &  (d\vecy_t - J \ev{\vecx_t}\,dt) \ev{\vecx_t}^T \approx \ev{(d\vecy_t - J \vecx_t) \vecx_{t}^T}, \label{eq:J Learning biol rog}
\end{eqnarray}
which takes the form of a local Hebbian learning rule.

\section*{Supporting Information}

\paragraph*{S1 Appendix.}
\label{S1}
{\bf The Neural Particle Filter: Mathematical Appendix.}

%\section*{Acknowledgments}
%We would like to thank Manfred Opper (TU Berlin) and Richard Hahnloser (INI, ETH and UZH) for fruitful discussions and Michael Pfeiffer (INI, ETH and UZH) for valuable feedback on the manuscript.

% Either type in your references using
% \begin{thebibliography}{}
% Text
% \end{thebibliography}
%
% or
%
% Compile your BiBTeX database using our plos2015.bst
% style file and paste the contents of your .bbl file
% here.
% 

\bibliographystyle{plos2015}

\end{document}

\maketitle

\section{Stochastic differential equations in a nutshell}

For readers who are not familiar with the concepts of It\^{o} stochastic differential equations (SDE), we want to give a very brief overview about how to describe diffusion processes, compute underlying probability distributions and moments from an SDE, and the common discretization scheme that we used to simulate the trajectories.

\subsection{Stochastic differential equations, moments and probability distributions}

Consider a vector-valued random variable $ \vecx_t \in \mathbb{R}^n$ that evolves according to an It\^{o} diffusion, i.e.~it can be described by the It\^{o} SDE
\begin{eqnarray}
d\mathbf{x}_{t} & = & \mathbf{a}(\mathbf{x}_{t},t)\, dt+ B(\mathbf{x}_{t},t) d\mathbf{w}_{t}.\label{eq:ItoSDE}
\end{eqnarray}
Here, $\mathbf{w}_{t}\in\mathbb{R}^{n}$ is a vector Brownian motion process with $\langle d\mathbf{w}_{t}d\mathbf{w}_{s}^{T}\rangle=\mathbb{I}^{n\times n}\delta_{ts}dt$, where $\mathbb{I}$ denotes the unit matrix in the corresponding dimension and further $\delta_{ts}=1$ if $ t=s $ and $\delta_{ts}=0$ otherwise.
The deterministic part of Eq.~\eqref{eq:ItoSDE} is determined by the \emph{drift term} $ \mathbf{a}(\mathbf{x}_{t},t) dt $ with a vector-valued function $ \mathbf{a}(\mathbf{x}_{t},t) $, whereas the stochastic part is determined by the \emph{diffusion term} $ B(\mathbf{x}_{t},t) d\mathbf{w}_{t} $ with the matrix-valued noise covariance $ B(\mathbf{x}_{t},t) $.

The process in Eq.~\eqref{eq:ItoSDE} defines a probability distribution $ p(\mathbf{x}_{t})$ over the random variable at each time $ t $.
In general, the evolution of the probability distribution $ p(\mathbf{x}_{t}) $ is given by the Fokker-Planck equation\footnote{For the sake of readability, we dropped the explicit $ t $-dependence in the arguments. This, however, does not affect the generality of Eq.~\eqref{eq:FokkerPlanckEquation} - \eqref{eq:FokkerPlanckOperator}.}: 
\begin{eqnarray}
dp(\mathbf{x}_t) & = & \mathcal{L}^\dagger\left[p(\mathbf{x}_t)\right]dt,\ \ \ \ \ \ \rm{with} \label{eq:FokkerPlanckEquation} \\
\mathcal{L}^\dagger\left[p(\mathbf{x}_t)\right] & = & - \sum_{i=1}^{n} \frac{\partial}{\partial x_{i}} \left[ a_{i}(\mathbf{x}_{t}) p(\mathbf{x}_t) \right] + \frac{1}{2} \sum_{i,j=1}^{n} \frac{\partial^{2}}{\partial x_{i} \partial x_{j}} \left[ B_{ij}(\mathbf{x}_t) p(\mathbf{x}_t) \right].
\label{eq:AdjointFokkerPlanckOperator}
\end{eqnarray}
where $ \mathcal{L}^\dagger $ is called adjoint Fokker-Planck operator \cite{Gardiner}.

%For a constant noise covariance $ \Sigma_{x} $, the choice of the nonlinear hidden dynamics $\mathbf{f}(\mathbf{x}_{t})$ will thus exclusively determine the shape of the probability distribution $ p(\mathbf{x}_{t}) $.
For certain drift and diffusion terms, there exists an analytical solution of Eq.~\eqref{eq:FokkerPlanckEquation} for the probability distribution $ p(\mathbf{x}_{t})$.
As an example, let us consider a stochastic process $ \mathbf{x}_{t} $ with drift $ a_{i}(\mathbf{x}) = a_{i}(x_{i}) $ and noise covariance $ B(\mathbf{x}_t)   = {\rm{diag} }(b_{i} (x_{i})) $.
For this process, the dimensions of the stochastic process are decoupled and the stationary distribution with $ dp(\mathbf{x}_{t})=0 $ can be computed from Eq.~\eqref{eq:FokkerPlanckEquation}:
\begin{eqnarray}
p(\mathbf{x}_{t})  & = &  \prod_{i=1}^{n} p(x_{t,i}) = \prod_{i=1}^{n} \frac{1}{Z_{i}}\exp\left(\int_{-\infty}^{x_{i}} \frac{a_i -\frac{1}{2} \frac{\partial}{\partial x'' } b_i (x'') \rvert_{x'} }{\frac{1}{2} b_i(x') } dx'\right), \label{eq:StationaryDistribution}
\end{eqnarray}
where $ Z_{i} $ denotes the normalization constant in dimension $ i $.\footnote{Note that depending on $ \mathbf{a}(\mathbf{x}) $ Eq.~\eqref{eq:StationaryDistribution} could for instance be non-normalizable, implying that no stationary distribution exists for this process.}

From the Fokker-Planck equation, it is possible to derive the evolution of the expectation of an arbitrary scalar-valued function $ \phi(\mathbf{x}) $:
\begin{eqnarray}
d\langle \phi  (\mathbf{x_t}) \rangle & = & \int d \mathbf{x}_t  \phi  (\mathbf{x}_t) \left( dp(\mathbf{x}_t)\right) \nonumber 
\\
& = & \left( \int d \mathbf{x}_t  \phi (\mathbf{x}_t) \mathcal{L}\left[p(\mathbf{x}_t)\right] \right) dt \nonumber
\\
& = &   \langle \sum_{i} a_i(\mathbf{x_t}) \frac{\partial}{\partial x_{t,i}} \phi (\mathbf{x_i}) + \frac{1}{2} \sum_{i,j}  B_{i,j}(\mathbf{x}_t) \frac{\partial^{2}}{\partial x_{t,i} \partial{x_{t,j}}} \phi(\mathbf{x}_t) \rangle dt \nonumber
\\
& =: & \langle \mathcal{L} \left( \phi (\mathbf{x_t}) \right) \rangle dt, \label{eq:dPhi}
\end{eqnarray}
where 
\begin{eqnarray}
\mathcal{L}\left[ \phi(\vecx_t) \right] & = & \sum_{i=1}^{n} a_{i}(\mathbf{x}_{t}) \frac{\partial}{\partial x_{i}}  \phi(\mathbf{x}_t) + \frac{1}{2} \sum_{i,j=1}^{n} B_{ij}(\mathbf{x}_t) \frac{\partial^{2}}{\partial x_{i} \partial x_{j}}   p(\mathbf{x}_t)
\label{eq:FokkerPlanckOperator}
\end{eqnarray}
$ \mathcal{L} $ is the Fokker-Planck operator.
To sum up, with a given It\^{o} SDE it is possible to set up equations for the evolution of its underlying probability distribution and for any scalar-valued function by determining the Fokker-Planck operator and its adjoint.

\subsection{Euler-Maruyama approximation}

For numerical simulation of the SDE in Eq.~\eqref{eq:ItoSDE} a time-discretization scheme, the so-called the Euler-Maruyama approximation, can be employed \cite{Kloeden1999} to integrate an It\^{o} SDE for small time steps $ \delta t = t_{n+1} - t_n $:
\begin{eqnarray}
\mathbf{x}_{n+1} & = & \mathbf{x}_{n} + \mathbf{a}(\mathbf{x}_{n},t_n) \delta t + B(\mathbf{x}_{n},t_n) \delta \mathbf{w}_n, \label{eq:EulerMaruyama}
\end{eqnarray}
where the increment of the Brownian motion process can be sampled from a Gaussian with zero mean and unit variance, i.e.~$ \delta \mathbf{w}_n \sim \mathcal{N}(0, \mathbb{I} \delta t) $.

The generative model in our manuscript (cf.~Eqs.~\eqref{eq:GenerativeModel Hidden} and \eqref{eq:GenerativeModel Observations}) directly defines the Gaussian transition probabilities $ p(\vecx_t | \vecx_{t-\delta t}) $ of the hidden state variable and the Gaussian emission probabilities $ p(\delta \vecy_t | \vecx_t)  $ for finite time steps $ \delta t $, which is called Euler-Maruyama approximation, \cite{Kloeden1999}:
\begin{eqnarray}
p(\vecx_t | \vecx_{t-\delta}) &=& \mathcal{N}\Big(\vecx_{t-\delta t} + \vecf(\vecx_{t-\delta t})\,\delta t, \Sigma_x \,\delta t \Big),  \label{eq:TransitionProbability}\\
p(\delta \vecy_t | \vecx_t ) &=& \mathcal{N} \Big( \vecg(\vecx_t)\,\delta t, \Sigma_y \, \delta t \Big) \label{eq:EmissionProbability}
\end{eqnarray}
The latter corresponds to the likelihood of the observations $ \delta \vecy_t $. 
Note that just because the transition and emission probabilities are Gaussian, the marginal probabilities $ p(\vecx_t) $ and $ p(\vecy_t) $ do not necessarily have to be Gaussians.
In fact, the shape of the marginal $ p(\vecx_t) $ is determined by the choice of the deterministic drift function $ \vecf(\vecx) $ and the noise covariance $ \Sigma_x $ such that it can be arbitrary.
This marginal then constitutes the prior over the real-world variable, which becomes time-invariant if the Fokker-Planck equation has a stationary solution $ p(\vecx_t,t) = p(\vecx)  $. 
The stationary solution of the Fokker-Planck equation for one-dimensional process and state-independent (additive) noise is given by
\begin{eqnarray}
p(x) &\propto& \exp\big( - 2 \frac{\phi(x) }{\sigma_x^2}  \big),
\end{eqnarray}
where $ \phi(x) = - \int_x f(x') dx' $ is the potential evoked by the deterministic drift.

\section{Nonlinear filtering in a nutshell}
%MAYBE move to SI

In this section, we will outline the formal solution to the filtering problem and introduce to approximate solution, which served as a benchmark in our manuscript: the standard particle filter and the extended Kalman filter.

\subsection{The formal solution to the filtering problem}

Recall that the general nonlinear filtering problem is based on the following generative model:
\begin{eqnarray}
d\vecx_t &=& \vecf(\vecx_t)\,dt+\Sigma_x^{1/2}\, d\mathbf{w}_t, \label{eq:GenerativeModel Hidden} \\
d\vecy_t &=& \vecg(\vecx)\,dt+\Sigma_y^{1/2}\,d\mathbf{v}_t. \label{eq:GenerativeModel Observations}
\end{eqnarray}

Solving the filtering problem for the generative model given by Eqs.~\eqref{eq:GenerativeModel Hidden} and \eqref{eq:GenerativeModel Observations} aims at computing the posterior probability of the hidden state $ p(\vecx_t | \Y_t) $, conditioned on the whole sequence of observations up to time $ t $.
This problem has already been recognized and tackled by mathematicians in the 60s and 70s of the last century, providing a formal solution for this problem in terms of a (stochastic) partial differential equation for the normalized posterior density, the so-called Kushner equation \cite{Kushner1964}
\begin{eqnarray}
d p(\vecx_t |Y_t ) &=& \FokkerPlanck^\dagger \big[ p \big] \, dt + \nonumber \\
& &  \Big( d\vecy_t - \ev{\vecg(\vecx_t)}_{p(\vecx_t |Y_t )} \Big)^T \Sigma_y^{-1} \Big( \vecg(\vecx_t) - \ev{\vecg(\vecx_t)}_{p(\vecx_t |Y_t )} \Big) p(\vecx_t |Y_t ), \label{eq:Kushner}
\end{eqnarray}
where $ \ev{\star} $ denotes the expectation with respect to the posterior probability\footnote{For the sake of readability, in the following we will drop the arguments in the functions $ \vecf(\vecx_t) $, $ \vecg(\vecx_t) $ and $ \phi(\vecx_t) $, and unless stated explicitly otherwise, expectations $ \ev{\star} $ are with respect to the posterior $ p(\vecx_t | \Y_t) $.} and $ \FokkerPlanck^\dagger[\star] $ denotes the adjoint Fokker-Planck operator of the hidden process in Eq.~\eqref{eq:GenerativeModel Hidden}.
The unnormalized posterior density $ \rho(\vecx_t | \Y_t) $ follows the so-called Zakai equation \cite{Wahrscheinlichkeitstheorie1969}
\begin{eqnarray}
d \rho(\vecx_t | \Y_t) & = & \FokkerPlanck^\dagger\big[ \rho \big] \, dt + \vecg^T \Sigma_y^{-1} d\vecy_t \, \rho(\vecx_t | \Y_t). \label{eq:Zakai}
\end{eqnarray}

Equivalently, the solution to the filtering problem is often given in terms of posterior expectations $ \ev{\phi(\vecx_t)} $ of an arbitrary function $ \phi(x) $ rather than in terms of the posterior density itself.
\begin{eqnarray}
d \ev{\phi} &=& \ev{\FokkerPlanck \big[ \phi \big]}\,dt + \text{cov}\big( \phi, \vecg^T \big) \Sigma_y^{-1} \big(d\vecy_t - \ev{\vecg}\,dt \big), \label{eq:KushnerEV}
\end{eqnarray}
where $ \phi(\vecx) $ is a twice-differentiable, scalar-valued function and  $ \FokkerPlanck $ is the Fokker-Planck operator for the hidden dynamics.
For example, $ \phi(\vecx) = \vecx $ determines the dynamics of the first moment of the posterior distribution, $ \vecmu_t = \ev{\vecx_t} $:
\begin{eqnarray}
d \vecmu_t &=& \ev{\vecf} \, dt + \text{cov}\big( \vecx_t, \vecg^T \big) \Sigma_y^{-1} \big(d\vecy_t - \ev{\vecg}\,dt \big). \label{eq:KushnerMean}
\end{eqnarray}

The formal solution to the filtering problem is, unfortunately, of little use for practical applications in most cases, because of the so-called closure problem.
To illustrate this problem, consider for instance the evolution of the first moment of one component of a one-dimensional hidden variable $ x $ for the generative function $ g(x) = x $, which we get by plugging $ \phi(x) = x $ into Eq.~\eqref{eq:KushnerEV}.
The prefactor of the second term in this equation, the covariance $ \text{cov} (\phi(x),g(x)) = \text{cov} (x,x) $ will then be a second-order moment.
In turn, if we want to compute the evolution of the second moment moment by setting $ \phi = x^2 $, we end up with a dependence on an even higher-order moment and so forth.
This problem has been recognized early on and makes these formal solutions infinite-dimensional.
Therefore, in order to solve the filtering problem in a general setting, we need to introduce suitable approximations.

\subsection{Approximations used as benchmark}

Here, we summarize the algorithms that were used as benchmark in the main manuscript.

\subsubsection{Particle Filter}

Particle filtering is a numerical technique to take samples from the posterior based on sequential importance sampling (see for instance \cite{Doucet2000} or \cite{Doucet2009} for review and thorough introduction) in general state-space models.
It is easily accessible, because in principle no knowledge of the Fokker-Planck equation or numerical methods for solving partial differential equations is needed. 
Here, we will briefly outline the algorithm, following \cite{Doucet2000}, and clarify how we used it with our generative model.

In general, one considers a state-space model of the form
\begin{eqnarray}
\vecx_t & \sim & p(\vecx_t|\vecx_{t-1} ), \\
\vecy_t & \sim  & p(\vecy_t|\vecx_{t}),
\end{eqnarray}
where $ \vecx_t $ denotes the hidden process following the prior transition probability $ p(\vecx_t|\vecx_{t-1} ), $ and $ \Y_t = \{\vecy_s, s\in \mathbb{N} \} $ is the sequence of observations generated from the emission probability $ p(\vecy_t|\vecx_{t}) $ at each time step.

Usually, samples cannot be taken from the filtering distribution $ p (\vecx_t | \Y_t) $ directly.
Instead, one takes $ N $ samples (commonly referred to as particles) $ \vecx_t^{(i)} $ from a proposal distribution $ \pi (\vecx_t | \X_{t-1}, \Y_t) $ conditioned on the whole history of observations $ \Y_t $ and all the previous states $ \X_{t-1} $.
These samples are weighted according to how well they correspond to the observations with a set of weights $ w_t^{(i)} $. 
The posterior is approximated by
\begin{eqnarray}
p(\vecx_t|\Y_t) & \approx & \sum_{i=1}^N w_t^{(i)} \delta(\vecx_t - \vecx_t^{(i)}) \\
\text{with}\ \ \ \ \sum_{i=1}^N w_t^{(i)} &=& 1.
\end{eqnarray}
In sequential importance sampling, sampling and reweighing is done recursively at each time step
\begin{eqnarray}
\vecx_t^{(i)} & \sim & \pi (\vecx_t | \X_{0:t-1}^{(i)}, \Y_t), \label{eq:PF proposal} \\
\tilde{w}_t^{(i)} & = & \tilde{w}_{t-1}^{(i)} \frac{p(\vecy_t|\vecx_t^{(i)})\, p(\vecx_t^{(i)}|\vecx_{t-1}^{(i)} ) }{\pi (\vecx_t^{(i)} | \X_{0:t-1}^{(i)}, \Y_t)}, \label{eq:PF weights unnormalized}\\
w_t^{(i)} & = & \frac{\tilde{w}_t^{(i)}}{\sum_{j=1}^{N} \tilde{w}_t^{(j)} }, \label{eq:PF weights}
\end{eqnarray}
where $ \tilde{w}_t^{(i)} $ denotes the unnormalized importance weight of particle $ i $ at time $ t $.

For our generative model in Eqs.~\eqref{eq:GenerativeModel Hidden} and \eqref{eq:GenerativeModel Observations}, choosing the prior transition probability as the proposal distribution, i.e.~$ \pi (\vecx_t | \X_{0:t-1}^{(i)}, \Y_t) = p(\vecx_t|\vecx_{t-1}^{(i)} ) $, we find that Eqs.~\eqref{eq:PF proposal} and \eqref{eq:PF weights unnormalized} are given by the emission and transition probability defined in Eqs.~\eqref{eq:TransitionProbability} and \eqref{eq:EmissionProbability}, respectively:
\begin{eqnarray}
\vecx_t^{(i)} & \sim & p(\vecx_t|\vecx_{t-1}^{(i)} ) = \mathcal{N}\Big(\vecx_{t-\delta t}^{(i)} + \vecf(\vecx_{t-\delta t}^{(i)})\, dt, \Sigma_x \, dt \Big), \\
\tilde{w}_t^{(i)} & = & \tilde{w}_{t-1}^{(i)} \, \mathcal{N} \Big( \vecg(\vecx_t^{(i)})\,dt, \Sigma_y \, dt \Big) .
\end{eqnarray}

This method has two disadvantages.
The first one is a problem called weight degeneracy:
After a finite number of iterations, all but one of the normalized importance weights are very close to zero, and only one particle will make a significant contribution to the posterior.
In our case, this is indeed catastrophic, because it implies that the posterior is approximated by a single independent sample from Eq.~\eqref{eq:GenerativeModel Hidden}, which does not take into account the observations at all.
To avoid this problem, we re-sample the particles in regular intervals with a probability proportional to their weight whenever the effective number of particles, 
\begin{eqnarray}
N_{\rm eff} = \frac{1}{\sum_j (w^{(j)})^2}
\end{eqnarray}
is smaller than $ N/3 $ and set their respective weights to $ 1/N $ accordingly.

The second shortcoming of this method is that it suffers from the curse of dimensionality, that is an exponential growth of computational complexity as a function of the dimension $ n $ of the state vector $ \vecx $.
To avoid this problem, more elaborate particle filters would have to be used, as is discussed elsewhere (e.g.~\cite{Rebeschini2015}).

\subsubsection{(Extended) Kalman filter}

In models where the hidden dynamics $ \vecf(\vecx) $ and the observation dynamics $ \vecg(\vecx) $ are linear, i.e.$ \vecf(\vecx) = A \vecx $ and  $ \vecg(\vecx) = J \vecx_t $, there exists a closed-form solution to Eq.~\eqref{eq:Kushner} corresponding to a Gaussian with time-varying mean $ \vecmu_t $ and variance $ \Sigma_t $.
The dynamics of these parameters are called the Kalman-Bucy filter \cite{Kalman1961} and form a set of coupled differential equations:
\begin{eqnarray}
d \vecmu_t &=& A \mu_t \, dt + \Sigma_t J^T \Sigma_y^{-1} \big( d\vecy_t - J \vecmu_t \, dt \big),\\
d \Sigma_t & =  & \Big( A \Sigma_t + \Sigma_t A^T + \Sigma_x - \Sigma_t J^T \Sigma_y^{-1} J \Sigma_t \Big)\, dt.
\end{eqnarray}
The dynamics of $ \Sigma_t $ is independent of the observations, consequently, the posterior variance becomes stationary after an initial transient.

In our work, as another model we compare our Neural Filter to we use is an approximation of the posterior for a nonlinear generative model called the extended Kalman filter.
It approximates the posterior by a Gaussian with mean $ \vecmu_t $ and variance $ \Sigma_t $ whose dynamics are derived by local linearization of the nonlinearities in the model.
This results in the following dynamics for the parameters:
\begin{eqnarray}
d \vecmu_t &=& \vecf(\vecmu)\, dt + \Sigma_y^{-1} \Sigma_t \big( d\vecy_t - \vecg(\vecmu_t) \, dt \big), \\
d \Sigma_t &=& \Big( F(\vecmu_t) \Sigma_t + \Sigma_t F(\vecmu_t)^T + \Sigma_x - \Sigma_t G(\vecmu_t)^T \Sigma_y^{-1} G(\vecmu_t) \Sigma_t \Big)\, dt,
\end{eqnarray}
where $ G_{ij} $
For models with multimodal posteriors, this approximation often breaks down, as we also see in our simulations.

\section{Log Likelihood Function}

\subsection{The likelihood function for stochastic processes}
Here, we want to outline the derivation of the log likelihood function by Moura \& Mitter \cite{Mourat1986}, which we use as a (negative) cost function for parameter learning.

Consider being given a sequence of observations $ \Y_t $ that may have been generated by two different diffusion processes, which, in turn, induce two different probability measures $ \mathbb{P} $ and $ \mathbb{Q} $, respectively.
In order to decide which of these processes is more likely to have generated the sequence, a likelihood ratio between these models can be computed, which corresponds to the so-called Radon-Nikodym derivative between the induced probability measures \cite[p.~282]{Klebaner2005}: 
\begin{eqnarray}
\Lambda (\mathcal{Y}_{t}) & = & \frac{d\mathbb{Q}(\mathcal{Y}_{t})}{d\mathbb{P}(\mathcal{Y}_{t})}. \label{eq:RadonNikodym}
\end{eqnarray}
Loosely speaking, a large value of $ \Lambda(\mathcal{Y}_{t}) $ provides evidence for the diffusion model inducing $ \mathbb{Q} $ and vice versa.

In our model, we have to consider that these observations $ \Y_t $ have been generated by a latent process $ \vecx_t $ following Eq.~\eqref{eq:GenerativeModel Hidden}.
Here we introduce the innovations process $ \vecn_t $, following the dynamics
\begin{eqnarray}
d \vecn_t = d \vecy_t - \ev{\vecg(\vecx_t)}_{p_\theta(\vecx_t|\Y_t)} dt. \label{eq:InnovationProcess}
\end{eqnarray}
The innovations process is a $ \Y_t $-adapted Brownian motion \cite[p.~33]{Bain2009} under the original measure $ \mathbb{Q}_\theta $ (or rather to a family of measures parametrized by $ \theta $), which is induced by Eqs.~\eqref{eq:GenerativeModel Hidden} and \eqref{eq:GenerativeModel Observations} with parameters $ \theta  $ and $ \Sigma_y=\mathbb{I} $.
By rearranging Eq.~\eqref{eq:InnovationProcess}, i.e.
\begin{eqnarray}
d \vecy_t =  \ev{\vecg(\vecx_t)}_{p(\vecx_t|\Y_t)} dt + d \vecn_t, \label{eq:InnovationY}
\end{eqnarray}
we thus obtain an It\^{o} SDE for the observations process $ \vecy_t $ under the original measure $ \mathbb{Q}_\theta(\mathcal{X},\mathcal{Y}) $, which additionally is completely independent of the latent process $ \vecx_t $, leaving us merely with a measure over the observations $ \mathbb{Q}_\theta(\mathcal{Y}) $.
The objective is now to maximize the likelihood that this process (Eq.~\ref{eq:InnovationY}) generated the observations with respect to the parameters $ \theta $.

In order to determine how likely the observations were generated from the model in Eq.~\eqref{eq:InnovationY}, we compute the Radon-Nikodym derivative of $ \mathbb{Q}_{\theta} $ with respect to the Wiener measure $ \mathbb{P} $, a measure under which $ \mathbf{y}_{t} $ is a Brownian motion process independent of hidden variables as well as parameters.\footnote{Here, the Wiener measure as a choice of reference is advantageous because the Radon-Nikodym derivative is straightforward to compute.}
The corresponding Radon-Nikodym derivative can be computed with Girsanov's theorem \cite[cf.~Eq.~3.18 on p.~52]{Bain2009}:
\begin{eqnarray}
\Lambda_\theta(\Y_t) &=& \frac{d\mathbb{Q}_\theta(\Y_t)}{d\mathbb{P}(\Y_t)} \\
&=& \exp \Big( \int_0^t \ev{\vecg(\vecx_t)}^T \, d\vecy_s - \frac{1}{2} \ev{\vecg(\vecx_s)}^T \ev{\vecg(\vecx_t)}\, ds \Big). \label{eq:LikelihoodRatio}
\end{eqnarray}
Equivalently, instead of maximizing this likelihood function, we consider the logarithm of the likelihood:
\begin{eqnarray}
L_t(\theta) &=& \ \int_0^t \ev{\vecg(\vecx_s)}^T \, d\vecy_s - \frac{1}{2} \ev{\vecg(\vecx_s)}^T \ev{\vecg(\vecx_s)}\, ds . \label{eq:LogLikelihoodRatio}
\end{eqnarray}

\subsection{An intuitive derivation of the likelihood ratio}

We want to give a less formal justification in discrete time and some intuition and further motivation for the objective function in Eq.~\eqref{eq:LikelihoodRatio}.

The overall objective for parameter learning is the maximization of the likelihood of the observation sequence $ \Y_t $.
Consider again the innovations process $ d \vecn_t $  (Eq.~\ref{eq:InnovationProcess}), which, conditioned on the whole sequence of observations $ \Y_t $ up to time $ t $, is a Brownian motion process.
In discrete time, at each (infinitely small) time step the increment $ d\vecy_t $ is thus distributed according to a Gaussian with mean $ \ev{\vecg(\vecx_t) }_{p(\vecx_t|\Y_t)} dt $ and variance $ dt $:
\begin{eqnarray}
p ( \Y_t ) &\approx& \prod_{s=0}^t p (d\vecy_s)  \\ % maybe a bit too sloppy? 
&\propto& \prod_{s=0}^t \exp \left( - \frac{(d\vecy_s - \ev{\vecg(\vecx_s)}dt)^2}{2 dt} \right) \\
\Rightarrow \log p ( \Y_t )	&=& - \sum_{s=0}^t \frac{(d\vecy_s - \ev{\vecg(\vecx_s)}dt)^2}{2 dt} + \text{const}(dt). \label{eq:ReconstructionError}
\end{eqnarray}
Considering again the continuous-time limit, the sum in $ \log p ( \Y_t )	 $ becomes an integral. 

To ensure Eq.~\eqref{eq:ReconstructionError} to be finite, any scaling with $ dt^m, m < 1 $ is undesirable.
However, $ d\vecy_t^2/dt$ is of order $ dt^0 $.
This term as well as the constant is eliminated by considering instead the logarithm of the likelihood \emph{ratio} between $ p(\Y_t) $ and $ q(\Y_t) $, where $ q(\Y_t) $ is the probability distribution of paths generated by a Brownian motion process with mean $ 0 $ and variance $ dt $
%Here, $ q(\Y_t) $ is the law in(? \textcolor{red}{not sure the formulation of this sentence makes sense)} the measure $ \mathbb{Q} $, under which $ \Y_t $ is a standard Brownian motion, i.e.~with mean $ 0 $ and variance $ dt $.
\begin{eqnarray}
L_t = \log \frac{p ( \Y_t )}{q(\Y_t)} &=& - \sum_{s=0}^t \frac{(d\vecy_s - \ev{\vecg(\vecx_s}dt)^2}{2 dt} + \sum_{s=0}^t \frac{d\vecy_s^2 }{2 dt} \nonumber \\
&\to& \int_0^t \ev{\vecg(\vecx_s)}^T d\vecy_s - \frac{1}{2} \ev{\vecg(\vecx_s)}^T \ev{\vecg(\vecx_s)} ds.
\end{eqnarray}
Note, that taking this ratio does not affect the maximization of Eq.~\eqref{eq:ReconstructionError}, because the denominator is independent of the model parameters as well as the hidden process $ \vecx_t $.

Another, and probably more intuitive, way to look at the objective is the following:
Equation \eqref{eq:ReconstructionError} actually corresponds to a \emph{prediction error}, i.e.~the difference between the actual observation $ d \vecy_t $ and its predicted value evaluated from the model parameters and the filtering distribution $ \ev{\vecg(\vecx_t) }_{p(\vecx_t|\Y_t)} dt $.

\subsection{Bias in the likelihood function for the particle estimate due to approximated posterior}
\label{section:Bias in the likelihood function}

In our model, the fact that we are using \emph{approximations} of the posterior instead of the real posterior  $ p(\vecx_t|\Y_t) $, which we don't have access to, in general introduces a bias into the posterior estimation of any function $ \ev{\phi(\vecx_t)} $.
In the following, we will use the notation $ \hat{\phi}_t = \mathbb{E}\big[\phi(\vecx_t)|\Y_t \big] $ for the \emph{true} posterior estimate (which we in general cannot calculate) and $ \ev{\phi(x)}_t \approx \frac{1}{N} \sum_{i=1}^{n} \phi(z_t^{(i)}) $ for the posterior estimate approximated by the particle positions.

In the general, nonlinear model, the bias $ \text{BIAS}(\ev{\phi}) = \widehat{\ev{\phi}-\phi} = \widehat{\ev{\phi}} - \hat{\phi}$ is not analytically accessible.
Moreover, this bias not only affects estimation, but also parameter learning, because the likelihood function also depends on posterior estimation and is a random variable even for offline learning\footnote{Note that for offline learning with the true posterior and for a fixed observation sequence, the log likelihood is deterministic.}.
Consequently, the gradient of the log likelihood that is used for parameter learning is biased as well.

Let us illustrate this for a simple, 1-dimensional generative model, which we attempt to solve with an SDE similar to the NPF and for which the bias in the log likelihood can be computed analytically:
\begin{eqnarray}
dx_t &=& ax_t\,dt + \sqrt{\Sigma_x} d \omega_t, \label{eq:LinearSystem_x}\\
dy_t &=& bx_t\,dt + \sqrt{\Sigma_y} d \nu_t.
\end{eqnarray}
The well-known optimal solution of this problem is a Gaussian with mean $ \mu_t = \hat{x}_t $ and variance $ \Sigma_t $, whose dynamics is given by the Kalman-Bucy filter \cite{Kalman1961}.
% \begin{eqnarray}
% d \mu &=& a \mu \,dt + \frac{\Sigma}{\Sigma_y}b(dy-b\mu\,dt), \label{eq:KBF_mu} \\
% d \Sigma &=& -\frac{b^2}{\Sigma_y}\Sigma^2\,dt + 2a\Sigma\,dt+\Sigma_x\,dt. \label{eq:KBF_Sigma}
% \end{eqnarray}
% Equation~\eqref{eq:KBF_Sigma} is independent of the innovations process and thus has a well-defined fixed point $ \Sigma^* $:
% \begin{eqnarray}
% \Sigma^* = \frac{a\Sigma_y}{b^2}+\frac{1}{b^2}\sqrt{\Sigma_y ( b^2\Sigma_x+a^2\Sigma_y ) }.
% \end{eqnarray}

Assuming stationary of the posterior, there exists a sampler of the form
\begin{eqnarray}
dz^{(i)}_t = \tilde{a} z_t^{(i)} \, dt + \tilde{b} \, dy_t + \tilde{c} \, dw_t,
\end{eqnarray}
which exactly resembles the posterior\footnote{unpublished work in our group, see also \cite{Greaves-Tunnell2015}}, because of the structural resemblance to the Kalman-Bucy filter (KBF).
Consequently, there exists a choice of the gain $ W_t $ in the equation of the NPF, such that the first moment of the posterior is matched with the first moment of the KBF.
We can compute the bias of the first moment of the single-particle estimator\footnote{The single-particle estimator denotes an estimator for the first moment of the posterior created by taking a single sample from the sampling equation.}:
\begin{eqnarray}
d z^{(i)}_t &=& a z^{(i)}_t \, dt + W (dy_t - bz_t^{(i)} \, dt ) + \sigma_x dw_t, \\
z\Rightarrow d \big( z^{(i)}_t - \hat{z}_t^{(i)} \big) &=& \big( a - W b \big) (z^{(i)}_t - \hat{z}_t^{(i)} ) \, dt + \sigma_x dw_t. \label{eq:Bias Mu of single particle posterior}
\end{eqnarray}
The first moment of the random process described by Eq.~\eqref{eq:Bias Mu of single particle posterior} is equal to the bias of the single-particle estimator, the second moment is equal to the variance of the estimator.
Since Eq.~\eqref{eq:Bias Mu of single particle posterior} is an Ornstein-Uhlenbeck process, the stationary solution is given by:
\begin{eqnarray}
\text{BIAS}\big( \ev{z^{(i)}} \big) &=& \lim_{t\to\inf} \widehat{ z^{(i)}_t - \hat{z}^{(i)}_t } = 0, \label{eq:Bias Bias of single particle posterior} \\
\text{VAR}\big( \ev{z^{(i)}} \big) &=& \lim_{t\to\inf} \widehat{ (z^{(i)}_t - \hat{z}^{(i)}_t)^2 } = \frac{\sigma_x^2}{2(Wb-a)}. \label{eq:Bias Variance of single particle posterior}
\end{eqnarray}

The bias in the single-particle estimator can be used to compute the bias of the log likelihood $ L $ for a limited number of samples.
The bias of the log likelihood is given by
\begin{eqnarray}
\text{BIAS}(\tilde{L})_t = \widehat{\tilde{L}}_t - L_t,
\end{eqnarray}
where
\begin{eqnarray}
L_t &=& \int_0^t b \hat{x}_s \, dy_s - \frac{1}{2} b^2 \hat{x}_s^2 ds , \\
\tilde{L}_t &=& \int_0^t b \ev{x}_s \, dy_s - \frac{1}{2} b^2 \ev{x}_s^2 ds \nonumber \\
	& 	=  &  \int_0^t \frac{b}{N}  \sum_i z_s^{(i)} \, dy_s - \frac{b^2}{2N^2} \sum_{i,j} z_s^{(i)} z_s^{(j)} ds,
\end{eqnarray}
are the true log likelihood and the log likelihood estimated from taking $ N $ samples, respectively.
With Eqs.~\eqref{eq:Bias Bias of single particle posterior} and \eqref{eq:Bias Variance of single particle posterior} it is straightforward to compute $ \widehat{\tilde{L}}_t $:
\begin{eqnarray}
\widehat{\tilde{L}}_t  &=& \int_0^t \frac{b}{N}  \sum_i \widehat{z}_s^{(i)} \, dy_s - \frac{b^2}{2N^2} \sum_{i,j} \Big( \frac{\sigma_x^2}{2(Wb-a)} \delta_{ij} + \widehat{z}_t^{(i)} \widehat{z}_t^{(j)}  \Big) ds \\
&=& L_t - \frac{1}{2} \frac{b^2}{N} \frac{\sigma_x^2}{2(Wb-a)} .
\end{eqnarray}
The second term denotes the bias of the estimated log likelihood.
This bias is always negative, i.e.~with a finite number of particles we systematically underestimate the true log likelihood, and asymptotically vanishes with increasing number of particles.

This example suggests that gradients for parameter learning should always be taken with respect to the bias-corrected log likelihood.
However, this bias cannot be estimated with a nonlinear model, because the true posterior estimated cannot be computed analytically.
Thus we have to accept that this bias introduces an error in our parameter estimation that is hard to control.
As long as we can be sure that our approximation is asymptotically correct, i.e.~in the limit of a large number of particles $ N \to \infty $ or an infinitely small binsize $ \delta_x \to 0 $ in the sampler, respectively, which is correct in the linear case, this bias vanishes asymptotically.

\section{Limits of large and small observation noise}
\label{section:si-limits-of-large-and-small-observation-noise}

We want to investigate the limits of small and large observation noise in the NPF and show that these limits are consistent with a Bayesian computation.
For the general nonlinear case, these cannot be calculated analytically, but it is possible to do so for a linear generative model, which we would like to outline in the following.

\subsection{Limits of the Kushner equation}

Before we start analysing the limits of the NPF, let us consider the ground-truth solution to the filtering problem once more: the Kushner equation (Eq.~\ref{eq:Kushner}.
Intuitively, for large observation noise the observations become meaningless and the Kushner equation becomes the Fokker-Planck equation, i.e.~the posterior evolves according to the prior.
Conversely, the Kushner equation should be guided entirely by the observation term in the limit of no observation noise, and for an invertible generative function $ \vecg(\vecx) $\footnote{A fact that we will assume in the following analysis.} the solution should become a delta function $ p (\vecx_t | \Y_t) \sim \delta (d\vecy_t - \vecg(\vecx_t) dt) $. %\textcolor{red}{I think g does not even need to be invertible if it is written like this.} 
It is tempting to say that of course this is the case, as the second term in Eq.~\eqref{eq:Kushner} is governed by $ \Sigma_y^{-1} $, meaning it should become large for $ \Sigma_y\to 0 $ and the Fokker-Planck term should be become negligible.
However, as $ \Sigma_y\to 0 $, so do $ \vecg-\ev{\vecg(\vecx)} $ and $ d\vecy - \ev{\vecg(\vecx)}dt $ and it is difficult to tell just from looking at the equation how fast these two contributions approach zero as $ \Sigma_y $ is reduced.

Let us therefore consider an illustrative example of a linear, 1-dimensional system\footnote{The number of dimensions does not affect the scaling with $ \Sigma_y $, which is why it readily generalizes to more dimensions}:
\begin{eqnarray}
dx &=& ax\,dt + \sqrt{\Sigma_x} dW, \label{eq:LinearSystem_x}\\
dy &=& bx\,dt + \sqrt{\Sigma_y} dV.
\end{eqnarray}
The well-known solution of this problem is a Gaussian with mean $ \mu $ and variance $ \Sigma $, whose dynamics is given by the Kalman-Bucy filter \cite{Kalman1961}:
\begin{eqnarray}
d \mu &=& a \mu \,dt + \frac{\Sigma}{\Sigma_y}b(dy-b\mu\,dt), \label{eq:KBF_mu} \\
d \Sigma &=& -\frac{b^2}{\Sigma_y}\Sigma^2\,dt + 2a\Sigma\,dt+\Sigma_x\,dt. \label{eq:KBF_Sigma}
\end{eqnarray}
Equation~\eqref{eq:KBF_Sigma} is independent of the innovations process and thus has a well-defined fixed point $ \Sigma^* $:
\begin{eqnarray}
\Sigma^* = \frac{a\Sigma_y}{b^2}+\frac{1}{b^2}\sqrt{\Sigma_y ( b^2\Sigma_x+a^2\Sigma_y ) }.
\end{eqnarray}

For $ \Sigma_y\to 0 $, we find that $ \Sigma^* $ approaches zero with $ \Sigma^* \propto \sqrt{\Sigma_y} $.
Using this relation, we find that the second term in Eq.~\eqref{eq:KBF_mu} scales with $ \sqrt{\Sigma_y}^{-1} $ and thus dominates the dynamics of $ \mu $.
Effectively, due to this very large prefactor the time scale of this dynamics $ \tau \propto \sqrt{\Sigma_y} $  approaches zero, such that $ \dot{y} $ can almost be seen as a constant, and $ \mu $ relaxes towards $ \mu = \dot{y}/b $ almost instantaneously.
So in the deterministic limit, the solution is indeed $ p(x_t|\Y_t) \sim \delta( dy - bx\,dt) $.

On the other hand, in the limit of large observation noise, we find $ \Sigma^*=-\frac{\Sigma_x}{2a} $, which corresponds to the variance of the prior probability distribution $ p(x) $ determined by Eq.~\eqref{eq:LinearSystem_x}.
Because it is independent of $ \Sigma_y $, the second term in Eq.~\eqref{eq:KBF_mu} vanishes and the dynamics of the mean are identical to that of the mean of Eq.~\eqref{eq:LinearSystem_x}.
Therefore, in the limit of large observation noise, the posterior distribution is identical to the prior probability distribution $  p(x_t|\Y_t) = p(x_t) $.

\subsection{Limits of the NPF}
For the empirical version of the NPF with $ W =  b \Sigma \Sigma_y^{-1} $, the analysis works analogous to the Kalman-Bucy case.
For our example, we find:
\begin{eqnarray}
d \mu &=& a \mu \,dt + \frac{\Sigma}{\Sigma_y}b(dy-b\mu\,dt), \label{eq:NFemp_mu} \\
d \Sigma &=& -\frac{2 b^2}{\Sigma_y}\Sigma^2\,dt + 2a\Sigma\,dt+\Sigma_x\,dt. \label{eq:NFemp_Sigma}
\end{eqnarray}
Note the factor of 2 that shows up in front of the quadratic term in Eq.~\eqref{eq:NFemp_Sigma}, which seriously affects the steady-state variance $ \Sigma^* $:
\begin{eqnarray}
\Sigma^* = \frac{a\Sigma_y}{2b^2}+\frac{1}{b^2}\sqrt{\Sigma_y ( 2b^2\Sigma_x+a^2\Sigma_y ) }.
\end{eqnarray}
However, it does not affect the solutions in the deterministic and zero-information limit, respectively, nor the proportionality with $ \sqrt{\Sigma_y} $ when letting $ \Sigma_y $ go to zero.
Thus, the same reasoning as in the previous sections applies and accordingly, the approximated posterior is the same as the real posterior.

\bibliographystyle{plos2015}